\documentclass[preprint]{ptephy_v2}%%%%%% to generate preprint number

\preprintnumber{XXXX-XXXX} %%% %%% Insert preprint number here
\usepackage{hyperref}
\usepackage[T1]{fontenc}
\usepackage{textcomp} % 欧文の補助記号（°など）
\usepackage{newtxtext,newtxmath}
\usepackage{bm,graphicx,placeins,subcaption}
\begin{document}

\title{Backward-angle electroproduction of $\eta '$ mesons off protons at $W=2.13~\mathrm{GeV}$ and $Q^{2}=0.46~\left(\mathrm{GeV/}c\right)^{2}$}

%%%% To generate auto affiliation numbers please use \author{}\affil{} command

\author[1,*]{T.~Akiyama}
\author[2]{P.~Byd\v{z}ovsk\'{y}}
\author[3]{T.~Gogami}
\author[4]{K.~Itabashi}
\author[5,6]{S.~Nagao}
\author[1,5,6]{S.~N.~Nakamura}
\author[1]{K.~Okuyama}
\author[7,8]{B.~Pandey}
\author[2]{D.~Skoupil}
\author[3]{K.~N.~Suzuki}
\author[7,9]{L.~Tang}
\author[10]{D.~Abrams}
\author[11]{D.~Androic}
\author[12]{K.~Aniol}
\author[13]{C.~Ayerbe~Gayoso}
\author[14]{J.~Bane}
\author[13]{S.~Barcus}
\author[14]{J.~Barrow}
\author[15]{V.~Bellini}
\author[16]{H.~Bhatt}
\author[16]{D.~Bhetuwal}
\author[7]{D.~Biswas}
\author[9]{A.~Camsonne}
\author[17]{J.~Castellanos}
\author[9]{J-P.~Chen}
\author[13]{J.~Chen}
\author[4]{S.~Covrig}
\author[18,19]{D.~Chrisman}
\author[20]{R.~Cruz-Torres}
\author[21]{R.~Das}
\author[22]{E.~Fuchey}
\author[10]{K.~Gnanvo}
\author[15,23]{F.~Garibaldi}
\author[7]{T.~Gautam}
\author[9]{J.~Gomez}
\author[7]{P.~Gueye}
\author[24]{T.~J.~Hague}
\author[9]{O.~Hansen}
\author[9]{W.~Henry}
\author[25]{F.~Hauenstein}
\author[9]{D.~W.~Higinbotham}
\author[54]{C.~E.~Hyde}
\author[1]{M.~Kaneta}
\author[9]{C.~Keppel}
\author[21]{T.~Kutz}
\author[7]{N.~Lashley-Colthirst}
\author[26,27]{S.~Li}
\author[28]{H.~Liu}
\author[28]{J.~Mammei}
\author[17]{P.~Markowitz}
\author[9]{R.~E.~McClellan}
\author[15,30]{F.~Meddi}
\author[9]{D.~Meekins}
\author[9]{R.~Michaels}
\author[31,32,33]{M.~Mihovilovi\v{c}}
\author[34]{A.~Moyer}
\author[20,35]{D.~Nguyen}
\author[24]{M.~Nycz}
\author[13]{V.~Owen}
\author[36]{C.~Palatchi}
\author[21]{S.~Park}
\author[11]{T.~Petkovic}
\author[10]{S.~Premathilake}
\author[37]{P.~E.~Reimer}
\author[17]{J.~Reinhold}
\author[37]{S.~Riordan}
\author[38]{V.~Rodriguez}
\author[39]{C.~Samanta}
\author[26]{S.~N.~Santiesteban}
\author[9]{B.~Sawatzky}
\author[31,32]{S.~\v{S}irca}
\author[26]{K.~Slifer}
\author[24]{T.~Su}
\author[40]{Y.~Tian}
\author[41]{Y.~Toyama}
\author[2]{D.~Trnkov\'{a}}
\author[1]{K.~Uehara}
\author[15]{G.~M.~Urciuoli}
\author[18,19]{D.~Votaw}
\author[42]{J.~Williamson}
\author[9]{B.~Wojtsekhowski}
\author[9]{S.~A.~Wood}
\author[26]{B.~Yale}
\author[37]{Z.~Ye}
\author[10]{J.~Zhang}
\author[10]{X.~Zheng}

\affil[1]{Department of Physics, Graduate School of Science, Tohoku University, Sendai, Miyagi 980-8578, Japan}%1
\affil[2]{Nuclear Physics Institute, CAS, \v{R}e\v{z}/Prague 25068, Czech Republic}%2
\affil[3]{Graduate School of Science, Kyoto University, Kyoto 606-8502, Japan}%3
\affil[4]{Department of Physics, Graduate School of Science, Osaka University, Toyonaka, Osaka 560-0043, Japan}%4
\affil[5]{Department of Physics, Graduate School of Science, The University of Tokyo, Hongo, Tokyo 113-0033, Japan}%5
\affil[6]{Quark Nucler Science Institute, Graduate School of Science, The University of Tokyo, Hongo, Tokyo 113-0033, Japan}%5
\affil[7]{Department of Physics, Hampton University, Hampton, Virginia 23668, USA}%6
\affil[8]{Department of Physics and Astronomy, Virginia Military Institute, Lexington, Virginia 24450, USA}%7
\affil[9]{Thomas Jefferson National Accelerator Facility, Newport News, Virginia 23606, USA}%8
\affil[10]{Department of Physics, University of Virginia, Charlottesville, Virginia 22904, USA}
\affil[11]{Department of Physics \& Department of Applied Physics, University of Zagreb, HR-10000 Zagreb, Croatia}
\affil[12]{Physics and Astronomy Department, California State University, Los Angeles, Califonia 90032, USA}
\affil[13]{Department of Physics, The College of William and Mary, Virginia 23185, USA}
\affil[14]{Department of Physics, University of Tennessee, Knoxville, Tennessee 37996, USA}
\affil[15]{INFN, Sezione di Roma, 00185, Rome, Italy}
\affil[16]{Department of Physics, Mississippi State University, Mississippi State, Mississippi 39762, USA}
\affil[17]{Department of Physics, Florida International University, Miami, Florida 33199, USA}
\affil[18]{Department of Physics and Astronomy, Michigan State University, East Lansing, Michigan 48824, USA}
\affil[19]{National Superconducting Cyclotron Laboratory, Michigan State University, East Lansing, MI 48824, USA}
\affil[20]{Department of Physics, Massachusetts Institute of Technology, Cambridge, Massachusetts 02139, USA}
\affil[21]{Department of Physics, State University of New York, Stony Brook, New York 11794, USA}
\affil[22]{Department of Physics, University of Connecticut, Storrs, Connecticut 06269, USA}
\affil[23]{Istituto Superiore di Sanit\`{a}, 00161, Rome, Italy}
\affil[24]{Department of Physics, Kent State University, Kent, Ohio 44242, USA}
\affil[25]{Department of Physics, Old Dominion University, Norfolk, Virginia 23529, USA}
\affil[26]{Department of Physics, University of New Hampshire, Durham, New Hampshire 03824, USA}
\affil[27]{Nuclear Science Division, Lawrence Berkeley National Laboratory, Berkeley, CA 94720, USA}
\affil[28]{Department of Physics, Columbia University, New York, New York 10027, USA}
\affil[29]{Department of Physics and Astronomy, University of Manitoba, Winnipeg, Manitoba R3T 2N2, Canada}
\affil[30]{Sapienza University of Rome, I-00185, Rome, Italy}
\affil[31]{Faculty of Mathematics and Physics, University of Ljubljana, 1000 Ljubljana, Slovenia}
\affil[32]{Jo\v{z}ef Stefan Institute, Ljubljana, Slovenia}
\affil[33]{Institut f\"{u}r Kernphysik, Johannes Gutenberg-Universit\"{a}t Mainz, DE-55128 Mainz, Germany}
\affil[34]{Department of Physics, Christopher Newport University, Newport News, Virginia 23606, USA}
\affil[35]{University of Education, Hue University, Hue City, Vietnam}
\affil[36]{Department of Physics, Indiana University, Indiana 47405, USA}
\affil[37]{Physics Division, Argonne National Laboratory, Lemont, Illinois 60439, USA}
\affil[38]{Divisi\'{o}n de Ciencias y Tecnologia, Universidad Ana G. M\'{e}ndez, Recinto de Cupey, San Juan 00926, Puerto Rico}
\affil[39]{Department of Physics \& Astronomy, Virginia Military Institute, Lexington, Virginia 24450, USA}
\affil[40]{Department of Physics, Syracuse University, New York, New York 10016, USA}
\affil[41]{Center for Muon Science and Technology, Chubu University, Aichi 487-8501, Japan}
\affil[42]{School of Physics \& Astronomy, University of Glasgow, Glasgow, G12 8QQ, Scotland, UK \email{takeru.akiyama.p5@dc.tohoku.ac.jp}}

%%% To include the collaborator name... Please use the command "\collaborator"
%%% For example: \collaborator{ATLAS Collaboration}

\begin{abstract}%
The electroproduction of $\eta '$ mesons from a $\mathrm{^{1}H}$ target at $W=\mathrm{2.13~GeV}$, $Q^{2} = 0.46~\left( \mathrm{GeV/}c\right)^{2}$ and $\cos \theta^{\mathrm{CM}}_{\gamma^{*}\eta '} \approx -1$ has been experimentally measured.
The differential cross section of virtual-photoproduction has been obtained as $4.4 \pm 0.8 ~\left( \mathrm{stat.} \right) \pm 0.4 ~\left( \mathrm{sys.} \right)~ \mathrm{nb/sr}$ in the One-Photon-Exchange Approximation.
This value is one-sixth of that of real-photoproduction at backward angles.
A comparison with newly-developed isobar model calculations not only shows validity of the theoretical framewark employed, but also imposes new constrains on coupling strength between the $\eta'p$ final state and nucleon resonances.
\end{abstract}

\subjectindex{xxxx, xxx}

\maketitle

\section{Introduction}
The quark model proposed by Gell-Mann~\cite{GM1,GM2,GM3} and Zweig~\cite{Zweig1, Zweig2} explains the fundamental properties of ground-state baryons and mesons based on the fravor $\mathrm{SU\left( 3\right)}$ symmetry with three lightest quarks ($u$, $d$ and $s$) and their antiquarks, but it does not fully account for the highly complex excited baryon spectra.
In particular, certain theoretically predicted but experimentally unestablished states are called ``missing resonances.''
Various meson photoproduction reactions have been extensively studied as powerful tool to explore such baryon resonances, owing to the fact that reactions mediated by electromagnetic interactions can be described with a high degree of theoretical precision.
Among the pseudoscalar meson nonet, the $\eta'$ meson has a mass of approximately $\mathrm{958~MeV/}c^{2}$ with isospin zero and no electric charge~\cite{PDG}.
This mass is significantly heavier than the predicted by the quark model due to quantum anomalies in QCD Lagrangian~\cite{ua1,ua2,ua3,ua4}.
In recent years, the changing of the mass of $\eta'$ mesons under high-density conditions has attracted considerable attention~\cite{EMN1,EMN2,EMN3,EMN4}, and experimental efforts have been undertaken to search for $\eta'$-mesic nuclei in order to explore the origin of the hadron's mass generation mechanism~\cite{TanakaGSI,TomidaLEPS}.
Also in this context, the cross section of the elementary production process is a fundamental quantity for these studies.
The photoproduction of $\eta '$ mesons off protons has three key characteristics: firstly, due to the heavy mass of $\eta '$, the reaction has a high energy threshold. Near this threshold, the available phase space is small and the reaction is dominated by low-order partial waves. Secondly, $\eta '$ mesons contains an $s\bar{s}$ quark-antiquark component, which may appear implicitly in the internal structure of the excited baryon resonances. Thirdly---and most importantly---the total isospin of the $\eta' p$ final state is $\frac{1}{2}$. Consequently, it can only couple to $N^{*}$ resonances with isospin $\frac{1}{2}$, meaning that decays from $\Delta$ resonances with isospin $\frac{3}{2}$ are forbidden.
The cross section of $\eta '$ photoproduction has been measured at various angles and $W$ settings by several collaborations such as SAPHIR~\cite{SAPHIR}, CLAS~\cite{CLAS06,CLAS09}, CBELSA/TAPS~\cite{CBELSA/TAPS}, A2MAMI~\cite{A2MAMI}, and LEPS~\cite{LEPS}. In particular, a detailed study of the cross section near the threshold has provided solid evidence for the existance of $N\left( 1895\right) \frac{1}{2}^{-}$ resonance.
However, there is a lack of data in the forward and backward angular regions in the center-of-mass (CM) frame relative to the photon beam, which contributes to the remaining model dependence in theoretical calculations describing the reaction. While experimental measurements in such extreme angular regions are difficult, they are expected to provide more sensitive information on resonance states with high angular momentum.\par

We have been conducting high-resolution missing mass spectroscopy of $\Lambda$ hypernuclei using an electron beam at the Thomas Jefferson National Accelerator Facility (JLab) for the past two decades~\cite{Miyoshi,Iodice,Cusanno,Nue,Tang,Urciuoli,Gogami}.  
In this series of experiments, a continuous primary electron beam is irradiated onto a target nucleus, producing $\Lambda$ hypernuclei via the $\left( e,e'K^{+}\right)$ reaction.
The forward-scattered electron and $K^{+}$ meson are measured using two magnetic spectrometers, and the mass spectrum of the $\Lambda$ hypernucleus is obtained as the missing mass.  
The most recent experiment, JLab E12-17-003, utilized a radioactive tritium gas target to search for the $nn\Lambda$ state, whose existence remains under debate~\cite{Suzuki,Pandey}.  
In our experiment, measuring $\Lambda / \Sigma^{0}$ production off protons is essential for energy calibration of the missing mass spectrum, while simultaneously providing differential cross sections for $\Lambda / \Sigma ^{0}$ electroproduction that can be compared with isobar model calculations describing $K^{+}$ meson photo- and electroproduction in a unified manner~\cite{Okuyama}.\par

This paper reports the first analysis of $\eta '$ production via the $\left( e,e'p\right)$ reaction within our experimental dataset.  
Since JLab E12-17-003 experiment did not apply any particle identification at the online level of data acquisition triggers, scattered protons can be identified in the hadron-side spectrometer using almost the same analysis method employed for $K^{+}$ identification.
Electroproduction, \textit{i.e.,} production reaction induced by electron scattering with non-zero four-momentum transfer $Q^{2}>0$, of $\eta '$ mesons have never been previously measured in experiments, nor has it been theoretically calculated. However, as with other meson production channels, the One-Photon-Exchange Approximation (OPEA) should hold in the low four-momentum transfer region (roughly $Q^{2}<1~\left(\mathrm{GeV/}c\right)^{2}$), allowing for a continuous interpretation in connection with real-photoproduction.  
In particular, it is noteworthy that our experimental setup, which is optimized for measurements of $\Lambda$ hypernuclei, also can observe $\eta '$ production at ultra-backward angles with respect to the virtual photon beam.
Additionally, we have developed a new isobar model calculation to describe $\eta 'p$ electroproduction by an extending BS model~\cite{BS12,BS3}, an isobar model calculation originally developed to describe the electroproduction of the $K^{+}\Lambda$ channel. This new calculation allows for comparisons between our experimental results of electroproduction and existing databases of photoproduction.
Through the comparison and interpretation, we aim to not only validate the theoretical framework of meson electroproduction, but also impose new constraints on the contributions of higher-mass nucleon resonances that couple to the $\eta 'p$ channel.\par

\section{Experiment}
\label{sec:expetiment}
\subsection{Principle}
\begin{figure}[t]
  \centering
  \includegraphics[width=0.7\linewidth]{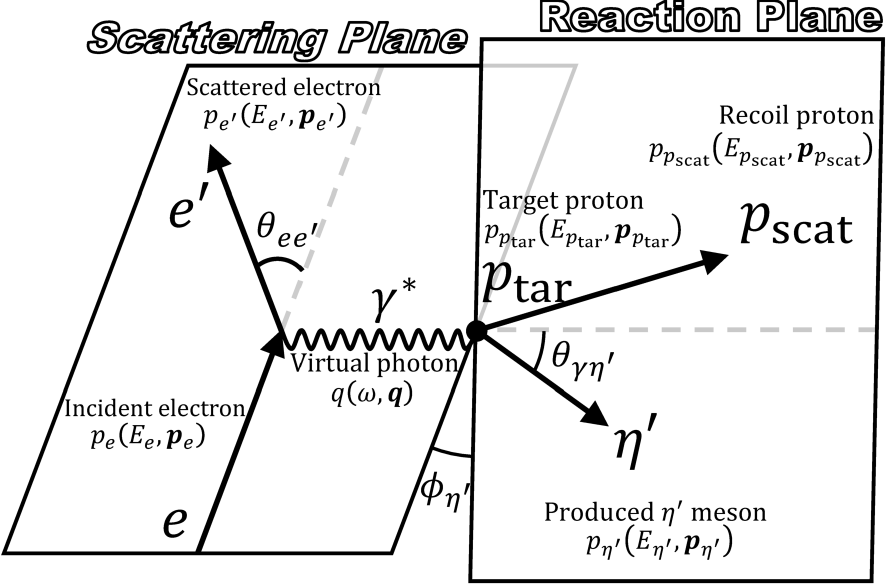}
  \caption{A schematic diagram based on OPEA of the $e+p \rightarrow e+p+\eta '$ reaction.}
  \label{reaction_diagram}
\end{figure}

The $\eta '$ mesons are produced with the $e+p \rightarrow e+p+\eta '$ reactions in the present measurement.
Fig.~\ref{reaction_diagram} shows a schematic diagram based on OPEA of the reaction.
A high-energy electron hits a fixed proton target and is scattered with some probability of virtual photon emission.
The virtual photon ($\gamma ^{*}$) with the energy and momentum $q = \left( \omega , \bm{q} \right) = \left( E_{e} - E_{e'}, \bm{p}_{e} - \bm{p}_{e'} \right)$ reacts with a target proton to produce an $\eta '$ meson.
The momenta of scattered electrons ($e'$) and recoil protons ($p_{\mathrm{scat}}$) are measured by two magnetic spectrometers.
The mass of the missing particle, \textit{i.e.} the missing mass, is obtained using the following equation deduced from the energy and momentum conservation,
\begin{equation}
  m_{\mathrm{X}} = \sqrt{ \left\{ \left( E_{e} - E_{e'} \right) + m_{p} - E_{p_{\mathrm{scat}}} \right\} ^{2} - { \left\{ \left( \bm{p}_{e} - \bm{p}_{e'} \right) - \bm{p}_{p_{\mathrm{scat}}} \right\} }^{2} }. \label{eq:MM}
\end{equation}
Events with the $\eta '$ productions are observed as a peak on the missing mass spectrum.
In OPEA, the triple differential cross section of the electroproduction of meson $\mathrm{X}$ is written as~\cite{MesonBook}
\begin{equation}
  \frac{\mathrm{d^{3}} \sigma}{\mathrm{d} E_{e'} \mathrm{d} \Omega _{e'} \mathrm{d} \Omega ^{\mathrm{CM}} _{\mathrm{X}}} = \Gamma \left( \frac{\mathrm{d} \sigma _{\gamma ^{*}}}{\mathrm{d} \Omega ^{\mathrm{CM}} _{\mathrm{X}}} \right) \label{eq:CS},
\end{equation}
where $\Gamma$ and $\frac{\mathrm{d} \sigma _{\gamma ^{*}}}{\mathrm{d} \Omega ^{\mathrm{CM}} _{\mathrm{X}}}$ are the flux of the virtual photons and the differential cross section in the center of mass (CM) frame of the hadron production by the virtual photon, respectively.
Conventionally, quantities for hadrons are described in the CM frame, whereas those for leptons are described in the laboratory frame.
An important benefit of OPEA is that, as in this expression, the triple differential cross section can be separated into two parts of electron scattering accompanied by virtual photon emission and the virtual photoproduction of mesons.
Then one can understand the reactions with virtual photons ($Q^{2} >0$) by analogy from the reactions with real photons ($Q^{2} = 0$) and compare them with each other.
Here, $Q^{2}$ is the four-momentum transfer with the electron scattering described as
\begin{equation}
 Q^{2} = -q^{2} = 2E_{e}E_{e'} - 2{m_{e}}^{2} - 2 \left| \bm{p}_{e} \right| \left| \bm{p}_{e'} \right| \cos \theta _{ee'}.
\end{equation}
It is worth to be noted that, in the present measurement, we measured the recoil protons instead of the generated $\eta '$ mesons.
The observable is therefore $\frac{\mathrm{d^{3}} \sigma}{\mathrm{d} E_{e'} \mathrm{d} \Omega _{e'} \mathrm{d} \Omega ^{\mathrm{CM}} _{p}}$ (or $\left( \frac{\mathrm{d} \sigma _{\gamma ^{*}}}{\mathrm{d} \Omega ^{\mathrm{CM}} _{p}} \right) _{\gamma ^{*} p \rightarrow p \eta '}$).
However, this cross section is identical to Eq.~(\ref{eq:CS}) since the recoil angle in a two-body reaction in the CM frame is back-to-back.
Assuming the ultra-relativistic approximation, the virtual photon flux $\Gamma$ is calculated by the following formula using the momentum of the incident and scattered electrons:
\begin{gather}
  \Gamma = \frac{\alpha}{2 \pi ^{2} Q^{2}} \frac{E_{\gamma}}{1-\epsilon} \frac{E_{e'}}{E_{e}} \label{eq:Gamma}\\
  \epsilon = \left[ 1 + 2 \frac{ \left| \bm{q} \right| ^{2}}{Q^{2}} \tan ^{2} \left( \frac{\theta _{ee'}}{2} \right) \right] ^{-1} \label{eq:epsilon}\\
 E_{\gamma} = \omega + \frac{q^{2}}{2m_{p}} \label{eq:Egamma},
\end{gather}
where $\epsilon$  denotes the transverse polarization of the virtual photons.
One can determine the total number of virtual photons by integrating $\Gamma$ with the total charge of the beam electrons and the acceptance of the electron spectrometer.
Then, the cross section of the $\gamma ^{*} + p \rightarrow \eta ' + p$ reaction, $\frac{\mathrm{d} \sigma _{\gamma ^{*}}}{\mathrm{d} \Omega ^{\mathrm{CM}} _{\mathrm{\eta '}}}$, is obtained.\par

The cross section for virtual photoproduction $\frac{\mathrm{d} \sigma _{\gamma ^{*}}}{\mathrm{d} \Omega ^{\mathrm{CM}} _{\mathrm{\eta '}}}$ can be expanded as follows depending on the polarization of the virtual photons and the azimuthal angle $\phi_{\eta '}$,
\begin{eqnarray}
  \frac{\mathrm{d} \sigma _{\gamma ^{*}}}{\mathrm{d} \Omega ^{\mathrm{CM}} _{\mathrm{\eta '}}} = \frac{\mathrm{d} \sigma _{\mathrm{T}}}{\mathrm{d} \Omega ^{\mathrm{CM}} _{\mathrm{\eta '}}} + \epsilon \frac{\mathrm{d} \sigma _{\mathrm{L}}}{\mathrm{d} \Omega ^{\mathrm{CM}} _{\mathrm{\eta '}}} + \sqrt{2 \epsilon \left( 1+\epsilon \right)} \frac{\mathrm{d} \sigma _{\mathrm{LT}}}{\mathrm{d} \Omega ^{\mathrm{CM}} _{\mathrm{\eta '}}} \cos \phi _{\eta '} + \epsilon \frac{\mathrm{d} \sigma _{\mathrm{TT}}}{\mathrm{d} \Omega ^{\mathrm{CM}} _{\mathrm{\eta '}}} \cos 2\phi _{\eta '} \label{eq:CS_henkyoku},
\end{eqnarray}
where, the particular contributions of $\frac{\mathrm{d} \sigma _{\mathrm{T}}}{\mathrm{d} \Omega ^{\mathrm{CM}} _{\mathrm{\eta '}}}$, $\frac{\mathrm{d} \sigma _{\mathrm{L}}}{\mathrm{d} \Omega ^{\mathrm{CM}} _{\mathrm{\eta '}}}$, $\frac{\mathrm{d} \sigma _{\mathrm{LT}}}{\mathrm{d} \Omega ^{\mathrm{CM}} _{\mathrm{\eta '}}}$, and $\frac{\mathrm{d} \sigma _{\mathrm{TT}}}{\mathrm{d} \Omega ^{\mathrm{CM}} _{\mathrm{\eta '}}}$ correspond to the transverse, longitudinal, transverse-longitudinal interference and transverse-transverse interference modes of the virtual photon~\cite{MesonBook}.
The terms ``L'' and ``LT'' appear only in reactions with virtual photons since real photons have only transverse wave components.
In the case $Q^2 \rightarrow 0$, only the transverse terms``T'' and ``TT'' remain, which correspond to the cross section and the polarized photon asymmetry in the photoproduction, respectively.\par

\subsection{Apparatus and setup}
\label{sec:setup}
The present experiment was conducted at JLab from October to November 2018 as JLab E12-17-003.
A continuous electron beam with an energy of $\mathrm{4.318~GeV}$ was supplied to Experimental Hall A by the Continuous Electron Beam Accelerator Facility (CEBAF).  
The beam energy uncertainty was approximately $1\times10^{-4}$ in full width at half maximum (FWHM)~\cite{CEBAF}.  
The typical beam current was $\mathrm{22.5~\mu A}$, and it was rastered to a size of $\mathrm{2.0\times2.0~mm^{2}}$ at the target position.  
These values for beam energy, current, and size were measured in real time during the beamtime period using the JLab Hall A beam monitoring system~\cite{arc,BCM,BPM}.  
We used a cryogenic gas target system enclosed in a cigar-shaped target cell~\cite{Target}.  
The gas targets available included $\mathrm{^{1}H}$, $\mathrm{^{2}H}$, $\mathrm{^{3}H}$, and $\mathrm{^{3}He}$, but in JLab E12-17-003, data were collected for $\mathrm{^{1}H}$ and $\mathrm{^{3}H}$.
We analyze the $\mathrm{^{1}H}$ data in this paper.
The target cell was made of aluminum alloy and had a length of $\mathrm{25~cm}$.  
The entire system was cooled to $\mathrm{40~K}$ by circulating liquid helium and using a copper heat sink.  
Additionally, a carbon multi-foil target for vertex position calibration was attached to the lower part of the target system.  
The total charge of the electron beam irradiated on the $\mathrm{^{1}H}$ target was $\mathrm{4.6~C}$, which corresponds to $\mathrm{2.9\times10^{19}}$~electrons.  
We used two High Resolution Spectrometers (HRS)~\cite{HallA}, which are the standard spectrometers of JLab Hall A, to measure scattered particles.
HRS-L measured the scattered electrons, while HRS-R measured the associated protons.
Each HRS has a QQDQ magnetic component, and their optical features are essentially identical.
The path length from the target position to the reference plane is $\mathrm{23.4~m}$.
The designed momentum resolution of a single HRS is $\mathrm{1\times10^{-4}}$ (FWHM), but in practice, the momentum resolution deteriorates than this value due to the finite material thickness (particularly of the target cell).
During the $\mathrm{^{1}H}$ target measurements, the central momenta were set to $p^{\mathrm{Lab}}_{e'} = \mathrm{2.100~GeV/}c$ for electrons and $p^{\mathrm{Lab}}_{p} = \mathrm{1.823~GeV/}c$ for protons.
The scattering angles of the two particles relative to the incident electron beam were set as $\theta^{\mathrm{Lab}}_{ee'} = -\theta^{\mathrm{Lab}}_{ep} = \mathrm{13.2~degrees}$.
Each HRS has a momentum acceptance of $\pm 4~\%$ relative to the central momentum and a solid angle coverage of approximately $\mathrm{5.5~msr}$.
These kinematic conditions were optimized to maximize the yield of $\Lambda$ hypernuclei, which is the primary objective of the present experiment.
From the view of the produced $\eta '$ mesons, this setup corresponds to a production angle of $\theta ^{\mathrm{CM}} _{\gamma^{*} \eta '} = \mathrm{172~degrees}$ (with $\mathrm{180~degrees}$ also within the acceptance range), relative to the virtual photon in CM frame.
We used the Vertical Drift Chambers (VDC1 and VDC2)~\cite{VDC} for particle tracking and the scintillation hodoscopes (S0 and S2) for time-of-flight measurements within HRS detector package.
Additionally, in HRS-R, data from two threshold-type aerogel Cherenkov detectors (AC1 and AC2; refractive indices $n = 1.015$ and $1.055$, respectively) were also taken for identification of hadrons. 
However, apart from energy calibration, these data were not used in the present analysis.
The data acquisition trigger was configured using a coincidence signal from the scintillation hodoscopes on both HRS-L and HRS-R.
It is noteworthy that neither of the two aerogel Cherenkov counters participated in the data acquisition trigger.
This configuration makes the measurement of $\eta '$ production, which requires identifying scattered protons instead of kaons, more straightforward.
A summary of the kinematic conditions for the present measurement is provided in Table~\ref{tb:kinematics_setting}.\par
\begin{table}[hbtp]
 \caption{Kinematical parameters of the present mesurement.}
 \label{tb:kinematics_setting}
 \centering
  \begin{tabular}{llc}
   \hline
   Item & Explanation& Value \\
   \hline \hline
   $E_{e}~\left[ \mathrm{GeV}\right]$ & Energy of electron beam ($e$) & $4.326$ \\
   $p_{e'}^{\mathrm{Lab}}~\left[ \mathrm{GeV/}c\right]$ & Momentum of scattered electron ($e'$) & $2.1$ \\
   $p_{p_{\mathrm{scat}}}^{\mathrm{Lab}}~\left[ \mathrm{GeV/}c\right]$ & Momentum of scattered proton ($p_{\mathrm{scat}}$) & $1.8$ \\
   $\theta^{\mathrm{Lab}}_{ee'}~\left[\mathrm{degrees}\right]$ & Zenith angle between $e$ and $e'$ & $13.2$ \\
   $\phi^{\mathrm{Lab}}_{ee'}~\left[\mathrm{degrees}\right]$ & Azimuthal angle between $e$ and $e'$ & $90$ \\
   $\theta^{\mathrm{Lab}}_{ep_{\mathrm{scat}}}~\left[\mathrm{degrees}\right]$ & Zenith angle between $e$ and $p_{\mathrm{scat}}$ & $13.2$ \\
   $\phi^{\mathrm{Lab}}_{ep_{\mathrm{scat}}}~\left[\mathrm{degrees}\right]$ & Azimuthal angle between $e$ and $p_{\mathrm{scat}}$ & $270$ \\
   \hline
   $\omega ~ \mathrm{\left[ GeV \right]}$ & Energy of virtual photon ($\gamma^{*}$) & $2.19$ \\
   $E_{\gamma} ~ \mathrm{\left[ GeV \right]}$ & Photon equivalent energy & $1.94$ \\
   $Q^{2} ~ \left[ \left( \mathrm{GeV/}c \right) ^{2} \right]$ & 4 momentum transfer & $0.467$ \\
   $W ~ \mathrm{\left[ GeV \right]}$ & Total energy of $\gamma^{*}+p_{\mathrm{tar}}$ in CM frame & $2.13$ \\
   $\theta ^{\mathrm{CM}} _{\gamma \eta '} ~ \mathrm{\left[ degrees \right]}$ & Zenith angle between $\gamma^{*}$ and $\eta'$ & $169$ \\
   $\phi ^{\mathrm{CM}} _{\gamma \eta '} ~ \mathrm{\left[ degrees \right]}$ & Azimuthal angle between $\gamma^{*}$ and $\eta'$ & $\left[-90, 90 \right]$ \\
   $\epsilon$ & Transverse polarization of $\gamma^{*}$ & $0.774$ \\
   $\left( \frac{d\sigma}{d\Omega} \right) _{\mathrm{CM}} / \left( \frac{d\sigma}{d\Omega} \right) _{\mathrm{Lab}}$ &Factor to transform Lab$\rightarrow$CM frame & $0.0679$ \\
   \hline
  \end{tabular}
\end{table}

\section{Analysis}
\subsection{Calibration}
The momentum, angle, and vertex position along the beam axis ($Z$-vertex) of scattered particles at the target position are reconstructed using hit information at the reference plane.  
The reconstruction are expressed in polynomial form, and the coefficients appearing in the formulas are optimized using calibration data.  
Details on this reconstruction and calibration method is given in Ref.~\cite{Suzuki} previously published by our collaboration.  
The position of the carbon multi-foil target was used for the calibration of the $Z$-vertex.  
The angle calibration was performed by referencing the hole pattern of the sieve slit installed at the entrance of both spectrometers.  
The momentum calibration of the scattered particles was performed by aligning the peaks in the missing mass of the $p\left(e,e'K^{+}\right)\Lambda/\Sigma^{0}$ reaction with the known values of masses of $m_{\Lambda} = \mathrm{1116~MeV/}c^{2}$ and $m_{\Sigma^{0}} = \mathrm{1193~MeV/}c^{2}$~\cite{PDG}.
During the momentum caliblation, data on the number of photoelectrons from the aerogel Cherenkov detector were utilized for $K^{+}$ identification.\par
\subsection{Event selection}
\begin{figure}[t]
  \centering
  \includegraphics[width=0.75\linewidth]{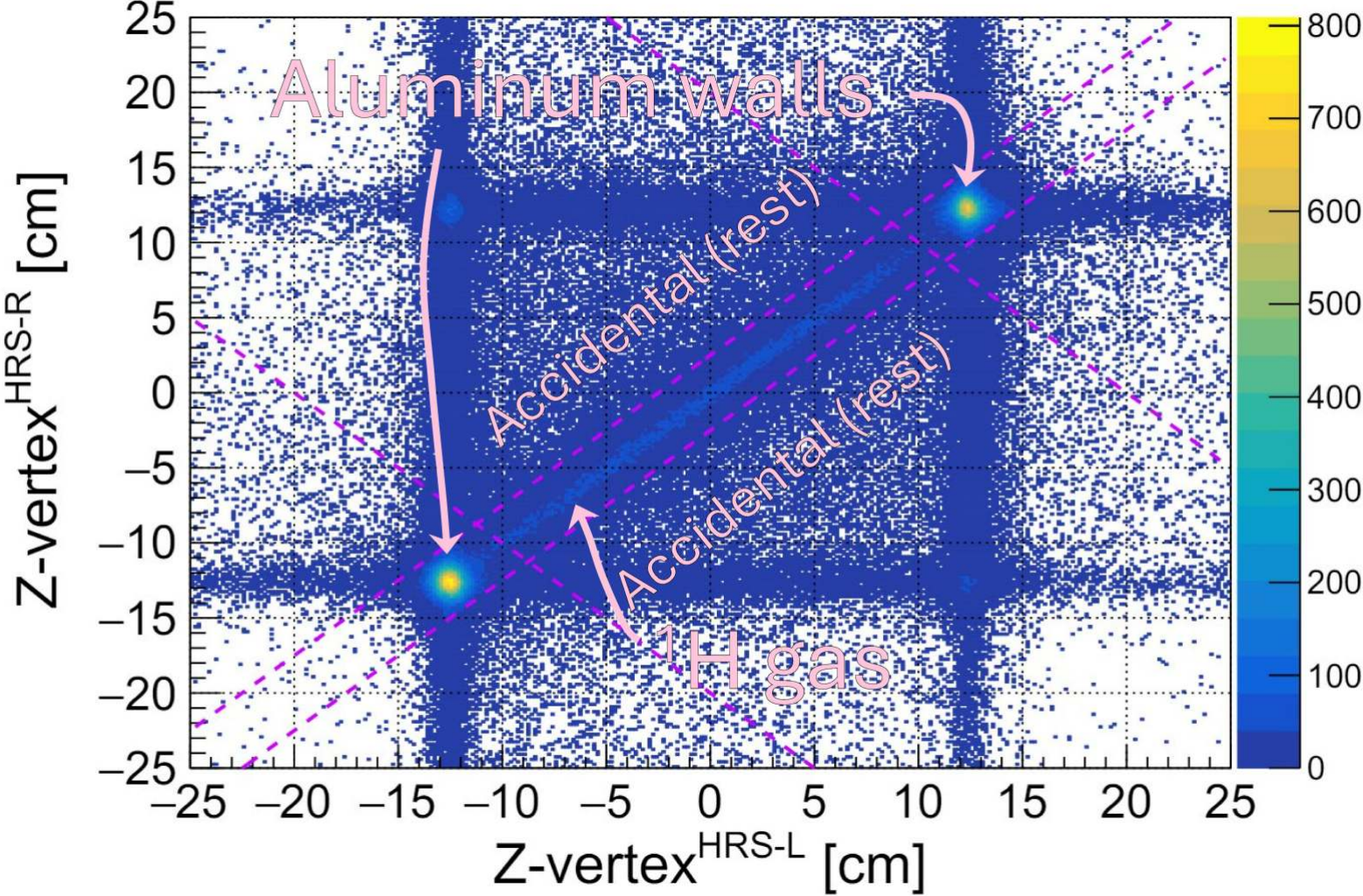}
  \caption{two-dimensional distribution of the $Z$-vertex reconstructed with both HRS-L and HRS-R. The regions of hydrogen gas and aluminum cells are clearly seen among the accidental background event. The violet dashed lines represent the gate of the target position selection for analysis.}
  \label{Z_LvsR}
\end{figure}

Applying appropriate event cuts to the $Z$-vertex distribution reconstructed by the left and right spectrometers helps to exclude events originating from reactions in the aluminum target cell, thereby selecting only those events corresponding to the $\mathrm{^{1}H}$ gas target region.
Fig.~\ref{Z_LvsR} shows the correlation of the $Z$-vertex reconstructed in both HRS-L and HRS-R.  
Note that each of the two spectrometers can independently reconstruct the $Z$-vertex.
In this graph, events located in the region with a positive correlation correspond to cases where both arms obtain consistent $Z$-vertex values.  
The events that are prominently concentrated around $Z = \mathrm{\pm15~cm}$ originate from reactions occurring in the aluminum walls located at the front and rear sides relative to the beam axis.  
Meanwhile, the region between these two peaks corresponds to the $\mathrm{^{1}H}$ gas target.  
On the other hand, events that are widely distributed outside these regions have inconsistencies in the reconstructed $Z$-vertex between the two arms.  
This region contains a significant number of events in which an electron and a hadron originating from unrelated reactions were detected in each arm.
These are referred to as ``accidental coincidence events''.  
To analyze events originating from the $\mathrm{^{1}H}$ gas target, we selected the region enclosed by the violet dashed lines in Fig.~\ref{Z_LvsR}.

\begin{figure}[t]
  \centering
  \includegraphics[width=0.75\linewidth]{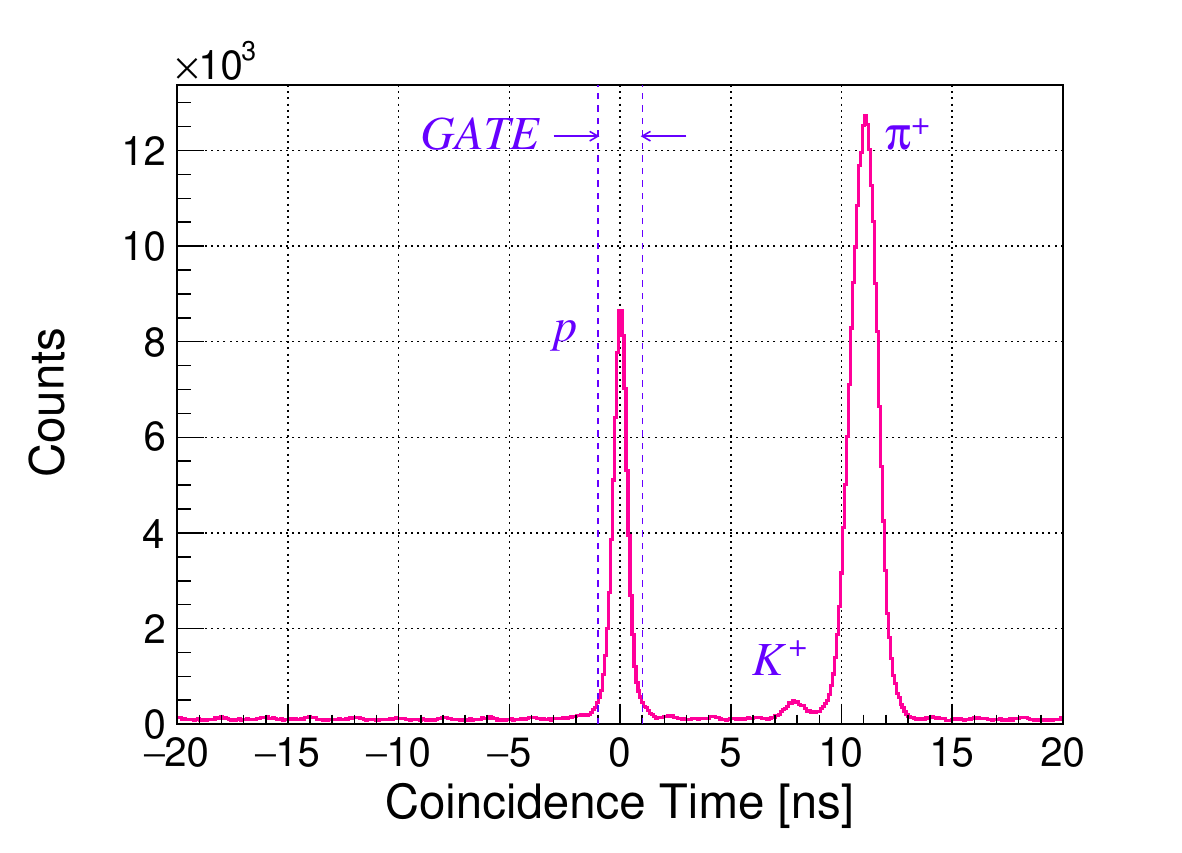}
  \caption{Coincidence time distribution. The peaks located at $\mathrm{0~ns}$, $\mathrm{\sim 7.8~ns}$, $\mathrm{\sim 11.1~ns}$ correspond to proton, kaon, and pion, respectively. Accidental coincidence events are distributed uniformly beneath the three peaks. The two blue dashed-lines represent the timing gate for proton selection.}
  \label{CoinTimeFit}
\end{figure}

The identification of hadrons in HRS-R was performed using the coincidence time ($T_{\mathrm{coin}}$) defined by the following equation:  
\begin{eqnarray}
  T_{ \mathrm{coin} } &:=& t_{\text{HRS-L}} \left( \mathrm{Target}\right) - t_{\text{HRS-R}} \left( \mathrm{Target}\right) \label{eq:cointime} \\
  t\left( \mathrm{Target} \right) &:=& t \left( \mathrm{Hodoscope}\right) - \frac{l_{\mathrm{path}}}{\beta c} \\
   &=& t \left( \mathrm{Hodoscope}\right) - \frac{\sqrt{p^{2}c^{2}+m^{2}c^{4}} \times l_{\mathrm{path}}}{p c^{2}}. \label{eq:timedif}
\end{eqnarray}
where $t_{\text{HRS-L}}\left(\mathrm{Target}\right)$ and $t_{\text{HRS-R}}\left(\mathrm{Target}\right)$ represent the reaction times at the target position, reconstructed from HRS-L and HRS-R, respectively.  
The reaction time at the target position is calculated using the hit time at hodoscopes $t \left( \mathrm{Hodoscope}\right)$ located downstream of the reference planes, the spectrometer's path length $l_{\mathrm{path}}$, and the particle's velocity $\beta$, which is obtained from momentum analysis.  
In HRS-R, $\beta$ is calculated under the assumption that the particle has the mass of a proton.  
Fig.~\ref{CoinTimeFit} shows the coincidence time distribution obtained from this analysis.  
A clear peak corresponding to proton events, which are correctly coincident at $\mathrm{0~ns}$, can be observed.  
On the other hand, other hadrons such as $\pi^{+}$ and $K^{+}$ form other peaks at different times due to the discrepancy in their assumed and actual masses.  
By applying the event cut represented by the blue dashed lines in Fig.~\ref{CoinTimeFit}, we selected approximately $4.8\times10^{4}$ proton events.\par

\subsection{Simulation}
\label{sec:sim}
We estimated the solid angles covered by HRS-L and HRS-R, as well as the shape of the missing mass spectrum consisting of the peak of $\eta'$ mesons and the background events originating from multi-meson production reactions, using a Monte Carlo simulation that models the setup of the present experiment, including the spectrometers and the target system.
This simulation was carried out using SIMC, a standard Monte Carlo simulation code developed and widely used for experiments in Hall A and Hall C at Jefferson Lab~\cite{SIMC1,SIMC2}.
SIMC calculates particle trajectories based on transport matrices that describe the magnetic features of the spectrometers, and it also simulates interactions between the beam particles and materials---such as energy loss due to ionization, multiple Coulomb scattering, and bremsstrahlung radiation---using Monte Carlo manner.
To properly account for the present experimental conditions, we newly modeled the aluminum target cell with a cigar-like shape~\cite{Target} and implemented it into SIMC.\par

\begin{figure}[t]
    \centering
    \begin{subfigure}[t]{0.45\textwidth}
        \centering
        \includegraphics[width=\linewidth]{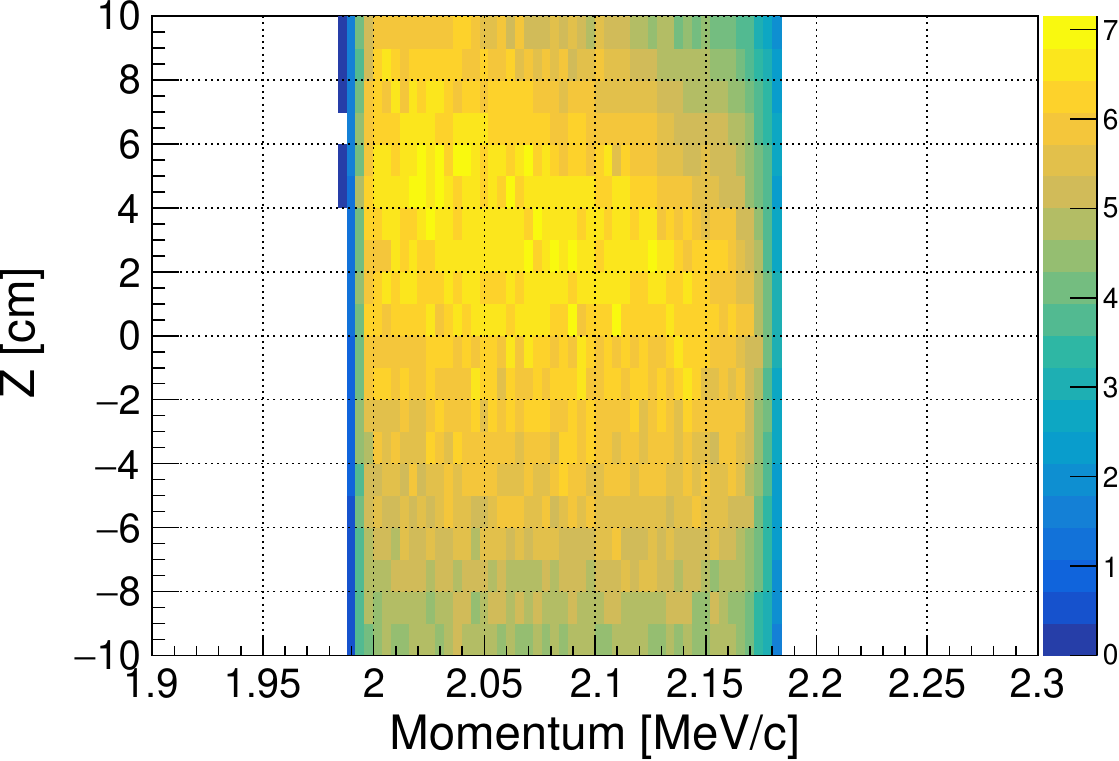}
        \caption{for HRS-L.}
        \label{sa_mom_z_L}
    \end{subfigure}
    \hspace{0.04\textwidth}
    \begin{subfigure}[t]{0.45\textwidth}
        \centering
        \includegraphics[width=\linewidth]{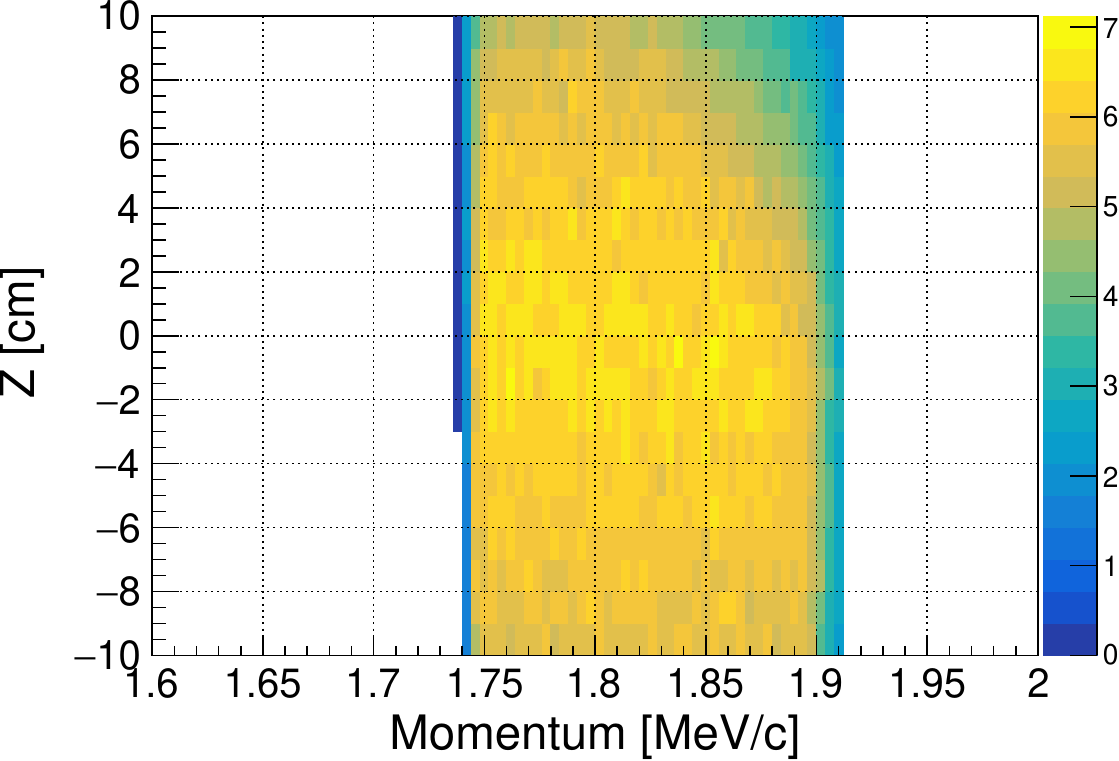}
        \caption{for HRS-R.}
        \label{sa_mom_z_R}
    \end{subfigure}
    \caption{The two-dimensional dependence of the solid angle on momentum and $Z$-vertex for HRS-L and HRS-R. The Z-axis represents the solid angle, with units of msr.}
    \label{sa_mom_z_LR}
\end{figure}

The solid angle of HRS is estimated through the following procedure: 1)~Particles are generated uniformly and randomly from the target position. The range of momentum and solid angle are set sufficiently wider than the acceptance region of the spectrometer, and the $Z$-vertex is distributed within the range of $\mathrm{\left[ -10~cm, 10~cm\right]}$.
2)~The solid angle acceptance of the spectrometer $\Delta\Omega_{\mathrm{acc}}$, is estimated using the following formula:
\begin{eqnarray}
  \Delta\Omega_{\mathrm{acc}} = \frac{N_{\mathrm{acc}}}{N_{\mathrm{gen}}}\Delta\Omega_{\mathrm{gen}},
\end{eqnarray}
where, $N_{\mathrm{gen}}$ is the total number of generated particles, $N_{\mathrm{acc}}$ is the number of particles that are transmitted through the spectrometer and counted as accepted events, and $\Delta\Omega_{\mathrm{gen}}$ is the solid angle over which particles are generated.
The momentum and $Z$-vertex dependence of the solid angles of HRS-L and HRS-R was estimated through simulations, as shown in Fig.~\ref{sa_mom_z_LR}.
This dependence is estimated as a two-dimensional solid angle map, and in the cross section calculation, the corresponding value is applied on an event-by-event basis.
The averaged solid angle over the full momentum and Z-vertex range is approximately 5.5 msr for both HRS-L and HRS-R.
Also on HRS-L side, the total number of virtual photons $N_{\gamma^{*}}$ irradiating the proton target is obtained by integrating the virtual photon flux $\Gamma$ as defined by Eqs.~(\ref{eq:Gamma})--(\ref{eq:Egamma}), over the number of incident electrons $N_{e}$, photon energy $\omega$, and the estimated solid angle $\Omega_{\text{HRS-L}}$:
\begin{eqnarray}
  N_{\gamma^{*}} = N_{e} \int \Gamma\mathrm{d}\omega\mathrm{d}\Omega_{\text{HRS-L}} .
\end{eqnarray}
Here, $N_{e}$ denotes the number of incident electrons that is obtained by dividing the total charge of $\mathrm{4.6~C}$ irradiated on the proton target (as mentioned in Sec.~\ref{sec:setup}) by the elementary charge $e = 1.6 \times 10^{-19}~\mathrm{C}$, yielding $N_{e} = 2.9 \times 10^{19}$.
The virtual photon flux $\Gamma$ is determined for each simulated event as based on the energy of the incident electron beam, the energy of the scattered electron, and its scattering angle.
As a result of the simulation, we estimate
\begin{eqnarray}
  N_{\gamma^{*}} = \left( 7.132 \pm 0.001~\left(\mathrm{stat}\right)\right) \times 10^{13},
\end{eqnarray}
which is used in the conversion to differential cross sections for virtual photoproduction.\par

\begin{figure}[t]
  \centering
  \includegraphics[width=12cm]{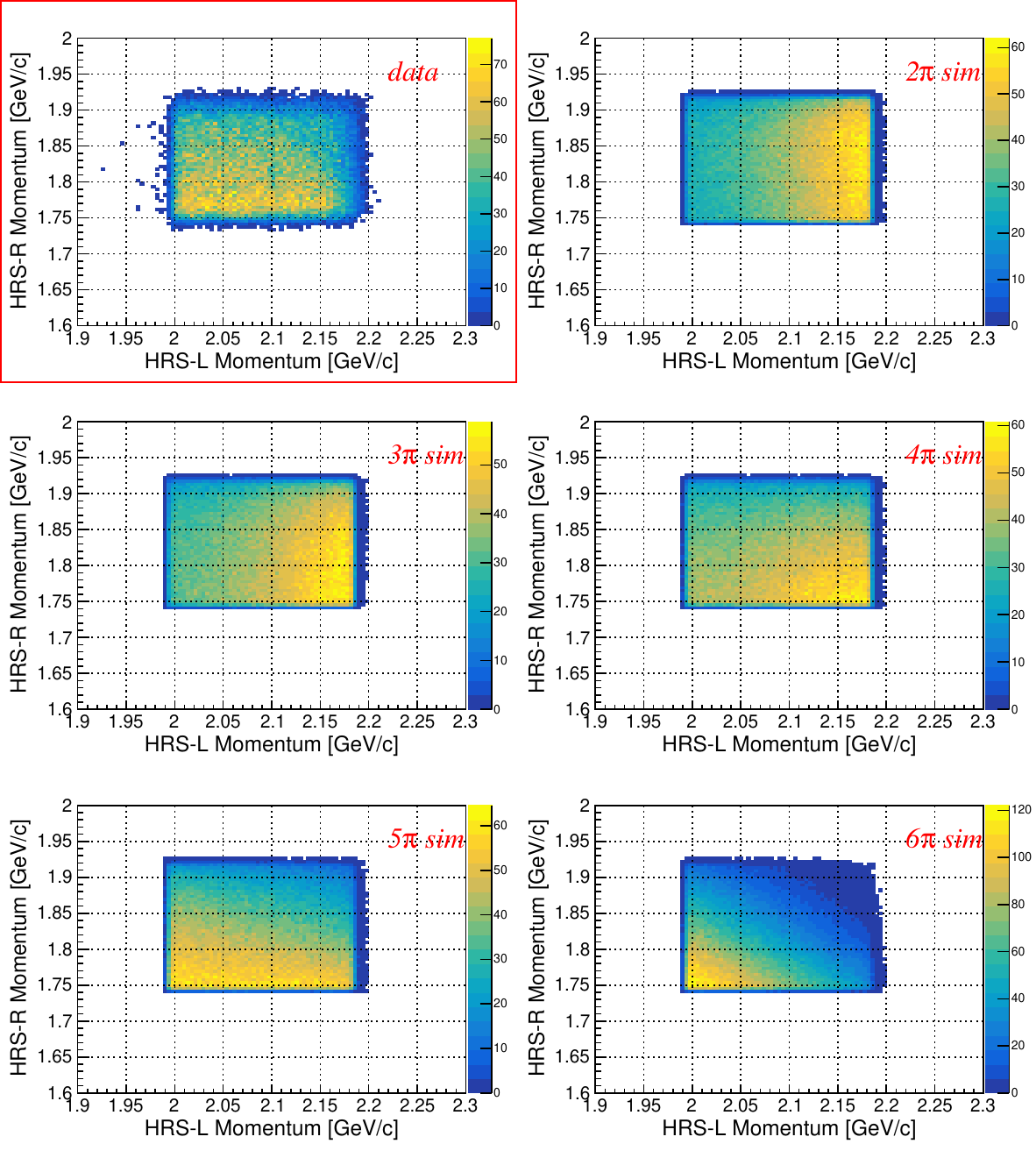}
  \caption{$p_{e'}$ v.s $p_{p_{\mathrm{scat}}}$ distribution for data and multi-pi simulation.}
  \label{moml_vs_momr_}
\end{figure}

We also estimated the shape of the missing mass spectrum under the present acceptance conditions.
The missing mass spectrum obtained in this experiment will be presented in Fig.~\ref{fig:combined2} in the following subsection, where a peak arising from $\eta'$ production along with a background component underneath it originating from multi-meson production is observed.
The shapes of both components in the spectrum are modeled through simulation, and the data fitting is performed using functional forms fixed accordingly.
First, we simulated the peak shape by generating electron–proton pairs according to the kinematics of $\eta'$ production.
The peak shape features a tail on the higher-mass side due to electron bremsstrahlung occurring in interactions with the aluminum walls of the target cell.
We then fitted the missing mass distribution with the following function composed of a superposition of two Gaussian functions sharing the same peak position, and an exponentially attenuating function smeared by those Gaussians:
\begin{eqnarray}
  F^{\mathrm{Peak}} \left( x \right) &:=& F^{\mathrm{DG}} \left(  x \right) + \int_{-\infty}^{+\infty} F^{\mathrm{DG}} \left(x - t\right) F^{\mathrm{Att}} \left( t \right) \mathrm{d}t \label{eq:peak_etap_1} \\
  F^{\mathrm{DG}} \left( x \right) &:=& A \left( 1 - C \right) \left[ R e^{-\frac{\left( x - \mu \right)^{2}}{2 \sigma_{1}^{2}}} + \left( 1 - R\right) e^{-\frac{\left( x - \mu \right)^{2}}{2 \sigma_{2}^{2}}} \right] \label{eq:peak_etap_2}\\
  F^{\mathrm{Att}} \left( x \right) &:=& \left\{
  \begin{array}{ll}
  0 & (x < \mu + \mu ')\\
  A C \tau e^{- \frac{x - \left( \mu' + \mu \right)}{\tau}} & (\mu + \mu ' \le x)
  \end{array}
  \right.
  . \label{eq:peak_etap_3}
\end{eqnarray}
Here, $F^{\mathrm{DG}}$ represents the peak component modeled by a double-Gaussian function, while $F^{\mathrm{Att}}$ denotes an asymmetric tail component described by the attenuation function.
This fitted function will be adopted as the function to be used in fitting the experimental data.
Next, the estimation of the background distribution in the missing mass spectrum is somewhat more complex.
We assumed that the background arises from multi-pion production and can be described as a superposition of events following the kinematics of non-resonant $2\pi\text{--}6\pi$ production.
Fig.~\ref{moml_vs_momr_} shows the distributions of $p_{e'}$ v.s $p_{p_{\mathrm{scat}}}$ obtained from the multi-pion production simulation, along with that obtained from the experimental data.
Each simulated distribution was scaled so that the number of events within the acceptance matches that of the experimental data.
When comparing simulations with different numbers of generated pions, we find that, as the number of pions increases, the momentum carried away by the pions becomes larger, thereby shifting the region where events are densely populated.
Comparison with the experimental data suggests that the distribution obtained for the case of five pions best reproduces the observed distribution.
In practice, the multiplicities of various pion production channels were linearly combined, and the consistency with the experimental data was evaluated using the $\chi^{2}$ value defined as:
\begin{eqnarray}
  \chi^{2} = \sum_{i: \mathrm{xbin}}\sum_{j: \mathrm{ybin}}\frac{\left(N_{i,j}^{\mathrm{data}} - N_{i,j}^{\mathrm{sim}}\right)^{2}}{\sigma_{i,j}^{2}}.
\end{eqnarray}
Here, $N_{i,j}^{\mathrm{data}}$, $N_{i,j}^{\mathrm{sim}}$, and $\sigma_{i,j}$ represent the number of events for the experimental data, the the number of events for the simulation, and the statistical uncertainty in the $\left(i,j\right)$-th bin, respectively.
By minimizing this $\chi^{2}$, we determined the set of weighting parameters that best reproduces the experimental data.
The result indicates that, without applying any selections on $Q^{2}$ or $W$, the 5-pion production channel is contributing 95.5~\% and the most dominant, while contributions from other channels were found to be negligibly small.
The average pion multiplicity was 4.86.
(Note that our interest here is solely in determining the functional form that best reproduces the experimental data. We do not intend to deduce the cross sections of each multi-pion production channels.)
By mapping the simulated events onto the missing mass axis, we determined the functional form to be used in fitting the missing mass spectrum of this experiment, which are presented in the next subsection.\par

\subsection{Missing mass}
\begin{figure}[t]
    \centering
    \begin{subfigure}{0.49\textwidth}
        \centering
        \includegraphics[width=\linewidth]{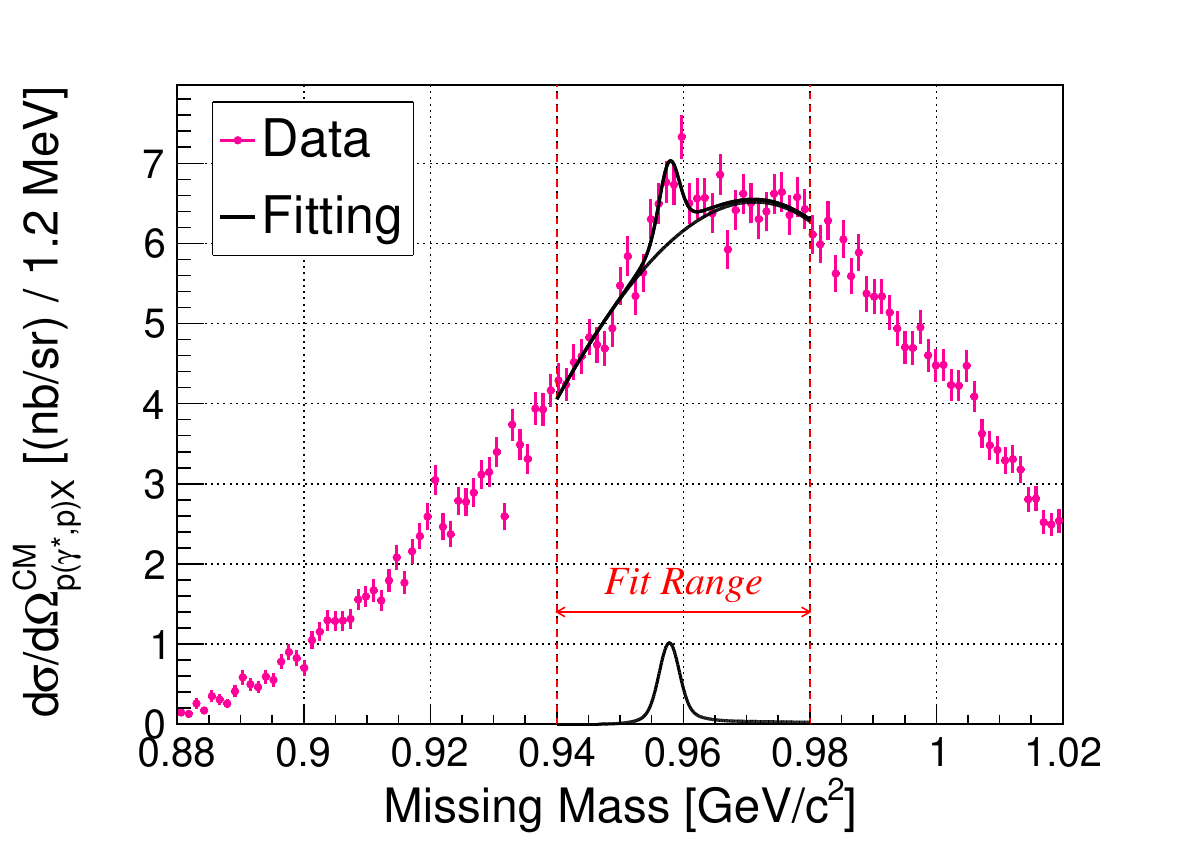}
        \caption{Using a cubic polynomial function for background.}
        \label{fig:subfig01}
    \end{subfigure}
    \hfill
    \begin{subfigure}{0.49\textwidth}
        \centering
        \includegraphics[width=\linewidth]{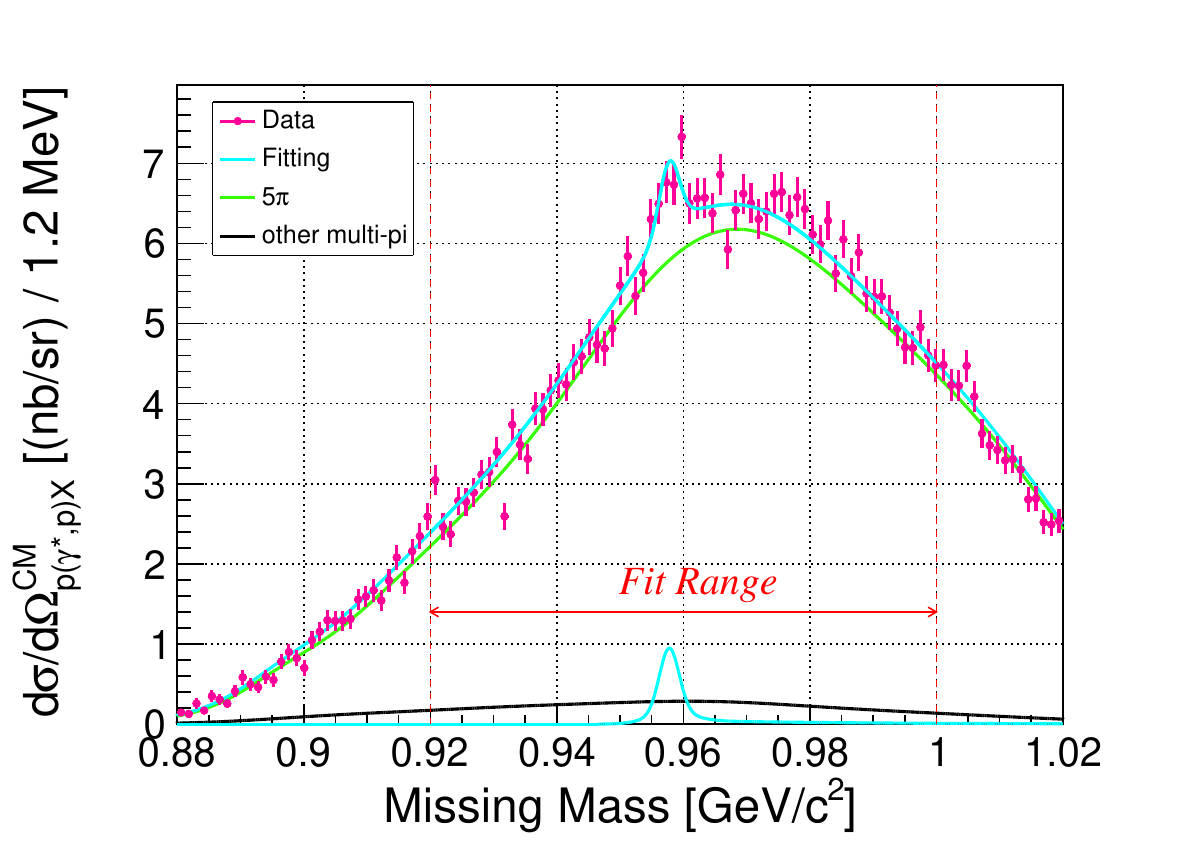}
        \caption{Using a simulation-based function for background.}
        \label{fig:subfig02}
    \end{subfigure}
    \caption{Missing mass spectrum (in differential cross section) of all data with different types of fitting function for background.}
    \label{fig:combined2}
\end{figure}

Fig.~\ref{fig:combined2} shows the missing mass spectrum of the $\mathrm{^{1}H}\left(e,e'p\right)X$ reaction, along with fittings using different functional forms for the background distribution.
In both panels (\ref{fig:subfig01}) and (\ref{fig:subfig02}), the data plots are identical.
On the broadly distributed background, an enhancement can be found near $m_{\eta '} = \mathrm{0.958~GeV/}c^{2}$.
It corresponds to events of the electroproduction of $\eta '$ mesons.  
The vertical axis has been converted into the differential cross section of the $\left(\gamma^{*},p\right)$ reaction, which is determined based on the OPEA as follows:  
\begin{gather}
 \left( \frac{\mathrm{d}\sigma _{\gamma^{*}p\rightarrow Xp}}{\mathrm{d}\Omega_{p}} \right) ^{\mathrm{CM}} = \frac{1}{N_{\mathrm{Target}}} \cdot \frac{1}{\varepsilon} \sum_{i=1}^{N_{\mathrm{Event}}} \frac{f^{\mathrm{Lab\rightarrow CM}}}{N_{\gamma ^{*}} \left( p_{e'}, z \right) \cdot \varepsilon_{i}^{\mathrm{DAQ}}  \cdot \Delta \Omega_{\text{HRS-R}} \left( p_{p}, z \right)} \label{eq:DC}\\
 \varepsilon = \varepsilon^{Z} \cdot \varepsilon^{\mathrm{CT}} \cdot \varepsilon^{\mathrm{RP}} \cdot \varepsilon^{\mathrm{Single}} \cdot \varepsilon^{\mathrm{Track}} \cdot \varepsilon^{\mathrm{Abs}} \cdot \varepsilon^{\mathrm{Detector}}.
\end{gather}
Here, $N_{\mathrm{Target}} = \mathrm{0.0375 \pm 0.00014 ~\left[ b^{-1} \right]}$ represents the number of target atoms, $N_{\gamma^{*}}$ is the number of virtual photon beams incident on the target, $\varepsilon = 0.474\pm0.006$ is the detection efficiency, $\Delta \Omega_{\text{HRS-R}}$ is the solid angle covered by HRS-R, and $f^{\mathrm{Lab\rightarrow CM}}$ is a factor to transform from the laboratory (Lab) frame to CM frame.  
$\Delta\Omega_{\text{HRS-R}}$, $N_{\gamma^{*}}$, and $f^{\mathrm{Lab\rightarrow CM}}$ were calculated on an event-by-event basis to be applied.
Also among various detection efficiencies, only the DAQ efficiency was measured run-by-run and applied separately for each run.
The descriptions and evaluated values of each efficiency factor are summarized in Table~\ref{tb:eff}.  
The differential cross section for $\eta'$ electroproduction under the kinematic conditions specified in this experiment is estimated by fitting the spectrum with an appropriate functional form.  
The functional form of the peak component was determined using simulations that generated electrons and protons following the kinematics of $\eta '$ production (as discussed in the previous subsection).
This function includes a tail component originating from bremsstrahlung in the target cell.  
On the other hand, the functional form of the background spectrum was assumed in two different ways.
The first method assumes a simple cubic polynomial, as shown in panel (\ref{fig:subfig01}) of Fig.~\ref{fig:combined2}.
In this case, the four coefficients of the polynomial are treated as free parameters in the fit.  
The second method determines the background functional form through a simulation (also see Sec.~\ref{sec:sim}), as shown in panel (\ref{fig:subfig02}) of Fig.~\ref{fig:combined2}.
During the fitting by this method, only the normalization (height) parameter of the background function was treated as a variable, while the other parameters that determine the shape of the background function were fixed.
The red dashed lines in both panels indicate the fit range.
Within the region, the reduced-$\chi^{2}$ values between the fit function and data were $1.07$ for panel (\ref{fig:subfig01}) and $1.49$ for (\ref{fig:subfig02}).  
The difference in the cross section obtained from the two fits was $11~\%$, which was incorporated as a systematic error due to the fitting method.
The next largest contribution to systematic error came from the uncertainty in estimating the spectrometer's solid angle.
By varying the momentum range used for analysis and recalculating the cross section, this uncertainty was evaluated to be $6~\%$.\par

\begin{table}[bt]
 \caption{Summary of each efficiency factor.}
 \label{tb:eff}
 \centering
  \begin{tabular}{llc}
   \hline
   Parameter & Explanation & Value \\
   \hline \hline
   $\varepsilon^{Z}$  & Efficiency for $Z$-vertex cut & $\mathrm{0.630}$\\
   $\varepsilon^{\mathrm{CT}}$  & Efficiency for coincidence time cut & $\mathrm{0.936}$\\
   $\varepsilon^{\mathrm{RP}}$  & Efficiency for hit position\&angle cut at reference plane& $\mathrm{0.886}$\\
   $\varepsilon^{\mathrm{Single}}$  & Efficiency for selection of single-track event& $\mathrm{0.970}$\\
   $\varepsilon^{\mathrm{Track}}$  & Tracking Efficiency in VDCs & $\mathrm{0.981}$\\
   $\varepsilon^{\mathrm{Abs}}$  & Particle's loss by absorption in target & $\mathrm{0.994}$\\
   $\varepsilon^{\mathrm{Detector}}$  & Detection efficiency in hodoscopes & $>\mathrm{0.999}$\\
   $\varepsilon^{\mathrm{DAQ}}$  & DAQ efficiency & $\mathrm{0.96~\text{(average)}}$\\
   \hline
  \end{tabular}
\end{table}

\section{Result and discussion}
\subsection{Result of the present measurement}
The differential cross section of the $\mathrm{^{1}H}\left( \gamma ^{*},p \right) \eta '$ reaction was obtained for the first time by fitting the missing mass spectrum:
\begin{eqnarray}
  \left( \frac{d\sigma _{\gamma^{*}p\rightarrow \eta 'p}}{d\Omega_{\eta '}} \right)  ^{\mathrm{CM}} = 4.4 \pm 0.8 ~\left( \mathrm{stat.} \right) \pm 0.4 ~\left( \mathrm{sys.} \right)~\left[ \mathrm{nb/sr}\right].
\end{eqnarray}
When the data is divided into two regions for $Q^{2}$ and $W$, the cross section is
\begin{eqnarray}
  \left( \frac{d\sigma _{\gamma^{*}p\rightarrow \eta 'p}}{d\Omega_{\eta '}} \right) ^{\mathrm{CM}} \left( Q^{2} < \left( 0.47~\mathrm{GeV/}c \right) ^{2} \right)= 5.9 \pm 1.1 ~\left( \mathrm{stat.} \right) \pm 0.6 ~\left( \mathrm{sys.} \right)~\left[ \mathrm{nb/sr} \right]  && \\
  \left( \frac{d\sigma _{\gamma^{*}p\rightarrow \eta 'p}}{d\Omega_{\eta '}} \right) ^{\mathrm{CM}} \left( Q^{2} \ge \left( 0.47~\mathrm{GeV/}c \right) ^{2} \right)= 4.3 \pm 1.1 ~\left( \mathrm{stat.} \right) \pm 0.6 ~\left( \mathrm{sys.} \right)~\left[ \mathrm{nb/sr} \right]  && \\
  \left( \frac{d\sigma _{\gamma^{*}p\rightarrow \eta 'p}}{d\Omega_{\eta '}} \right) ^{\mathrm{CM}} \left( W < 2.13~\mathrm{GeV} \right)=  3.7 \pm 1.2 ~\left( \mathrm{stat.} \right) \pm 0.7 ~\left( \mathrm{sys.} \right) \hspace{25pt}~\left[ \mathrm{nb/sr} \right]  && \\
  \left( \frac{d\sigma _{\gamma^{*}p\rightarrow \eta 'p}}{d\Omega_{\eta '}} \right) ^{\mathrm{CM}} \left( W \ge 2.13~\mathrm{GeV} \right)= 6.5 \pm 1.0 ~\left( \mathrm{stat.} \right) \pm 0.8 ~\left( \mathrm{sys.} \right) \hspace{25pt}~\left[ \mathrm{nb/sr} \right]  &&. 
\end{eqnarray}
\par

\subsection{Decomposition of the differential cross section}
As given at Eq.~(\ref{eq:CS_henkyoku}), the differential cross section of meson electroproduction is decomposed into four terms of $\frac{\mathrm{d} \sigma _{\mathrm{T}}}{\mathrm{d} \Omega ^{\mathrm{CM}} _{\mathrm{\eta '}}}$, $\frac{\sigma _{\mathrm{L}}}{\mathrm{d} \Omega ^{\mathrm{CM}} _{\mathrm{\eta '}}}$, $\frac{\mathrm{d} \sigma _{\mathrm{LT}}}{\mathrm{d} \Omega ^{\mathrm{CM}} _{\mathrm{\eta '}}}$, and $\frac{\mathrm{d} \sigma _{\mathrm{TT}}}{\mathrm{d} \Omega ^{\mathrm{CM}} _{\mathrm{\eta '}}}$.
The experimental setup for the present measurement has an almost constant acceptance of $-90~\left[ \mathrm{degrees}\right] < \phi_{\eta '} < 90~\left[ \mathrm{degrees}\right]$ for the angle $ \phi_{\eta '}$ between the reaction plane and the scattering plane (also see Fig.~\ref{reaction_diagram}).
Averaging Eq.~(\ref{eq:CS_henkyoku}) within this $ \phi_{\eta '}$ range cancels the $\frac{\mathrm{d} \sigma _{\mathrm{TT}}}{\mathrm{d} \Omega ^{\mathrm{CM}} _{\mathrm{\eta '}}}$ term and brings
\begin{eqnarray}
  \left( \frac{\mathrm{d} \sigma _{\gamma ^{*}}}{\mathrm{d} \Omega ^{\mathrm{CM}} _{\mathrm{\eta '}}} \right) _{\mathrm{ave.}} = \frac{\mathrm{d} \sigma _{\mathrm{T}}}{\mathrm{d} \Omega ^{\mathrm{CM}} _{\mathrm{\eta '}}} + \epsilon \frac{\mathrm{d} \sigma _{\mathrm{L}}}{\mathrm{d} \Omega ^{\mathrm{CM}} _{\mathrm{\eta '}}} + \frac{2}{\pi} \sqrt{2 \epsilon \left( 1+\epsilon \right)} \frac{\mathrm{d} \sigma _{\mathrm{LT}}}{\mathrm{d} \Omega ^{\mathrm{CM}} _{\mathrm{\eta '}}} \label{eq:CS_decomp}.
\end{eqnarray}
This $\left( \frac{\mathrm{d} \sigma _{\gamma ^{*}}}{\mathrm{d} \Omega ^{\mathrm{CM}} _{\mathrm{\eta '}}} \right) _{\mathrm{ave.}}$ corresponds to the observable in this experiment.
In calculations introduced in the next subsection, it is possible to directly calculate $\frac{\mathrm{d} \sigma _{\mathrm{T}}}{\mathrm{d} \Omega ^{\mathrm{CM}} _{\mathrm{\eta '}}}$, $\frac{\mathrm{d} \sigma _{\mathrm{L}}}{\mathrm{d} \Omega ^{\mathrm{CM}} _{\mathrm{\eta '}}}$, $\frac{\mathrm{d} \sigma _{\mathrm{LT}}}{\mathrm{d} \Omega ^{\mathrm{CM}} _{\mathrm{\eta '}}}$, and $\frac{\mathrm{d} \sigma _{\mathrm{TT}}}{\mathrm{d} \Omega ^{\mathrm{CM}} _{\mathrm{\eta '}}}$ in the specific kinematics.
Therefore we compose $\left( \frac{\mathrm{d} \sigma _{\gamma ^{*}}}{\mathrm{d} \Omega ^{\mathrm{CM}} _{\mathrm{\eta '}}} \right) _{\mathrm{ave.}}$ from particular terms.\par

\subsection{Isobar model to the $\eta'$ electroproduction}
We demonstrate a newly-developed isobar model calculation for describing the photo- and electro-production of $\eta'$ mesons.
The new calculation was constructed by extending the framework of BS model~\cite{BS12,BS3}, which has been successfully established in the $K^{+}\Lambda$ channel, to the $\eta'N$ channel based on OPEA.
The development of this new calculation is detailed in Ref.~\cite{Skoupil} by the members of our author team responsible for the theoretical part.\par

The isobar model is one of the widely-accepted theoretical models used to describe meson production reactions, and it takes into account effective Lagrangians with hadronic degrees of freedom.
In the isobar models, only the lowest-order perturbative terms---referred to as ``tree-level'' contributions--- in the $s$-, $t$-, and $u$-channels are considered, and then the effective coupling constants at each vertex are determined by fitting to the experimental database.
Intermediate states are assumed to include not only the ground state nucleon but also various excited nucleon resonances.
As a result, quarks itself do not explicitly appear in the isobar model; instead, complex features of the reaction dynamics are modeled by the number and various types of resonances included in the assumed set.\par

Different types of form factors in virtual-photoproduction are introduced to account for the spatial distribution of hadrons.
First, The electromagnetic form factor based on the Extended Vector Meson Dominance (EVMD) model~\cite{EVMD1,EVMD2} is applied at the virtual-photocoupling vertex.
In EVMD framework, the coupling between the photon and the hadron is described as the sum of the direct coupling term to quarks inside the hadron and terms involving explicit couplings to vector mesons ($\rho^{0}$, $\omega$ and $\phi$) that share the same quantum numbers as the photon.
The electromagnetic form factor has an explicit dipole-like dependence on four momentum transfer $Q^{2}$, and directly affects the $Q^{2}$-dependence of the differential cross section in meson electroproduction.
In the new calculation, we adopt the GKex(02S) model by E.~L.~Lomon's parametrization~\cite{EMFF2}, an update of that by R.~A.~Williams~\textit{et.~al.}~\cite{EMFF1} for the $s$- and $u$-channel baryon exchanges.
Meanwhile, the hadronic form factor is introduced at the hadronic-coupling vertex and, in this study, is expressed by a simple dipole function with respect to the Mandelstam variables $x = s, t$ and $u$:
\begin{eqnarray}
  F_{\mathrm{h.f.f.}} = \left[ \frac{\Lambda^{4}}{\left( x- M_{R}^{2} \right) ^{2} + \Lambda^{4}} \right] ^{1/2},
\end{eqnarray}
or, a multi-dipole function:
\begin{eqnarray}
  F_{\mathrm{h.f.f.}} = \left[ \frac{\Lambda^{4}}{\left( x- M_{R}^{2} \right) ^{2} + \Lambda^{4}} \right] ^{J_{R}+1/2},
\end{eqnarray}
where, $J_{R}$ denotes the spin of the resonance~\cite{BS12}.
The inclusion of the hadronic form factor is essential for ensuring consistency with the photoproduction database.
The cutoff parameters $\Lambda$ for the resonances and background channels are also determined by fitting to the experimental photoproduction data.\par

\begin{figure}[btp]
  \centering
  \includegraphics[width=14.0cm]{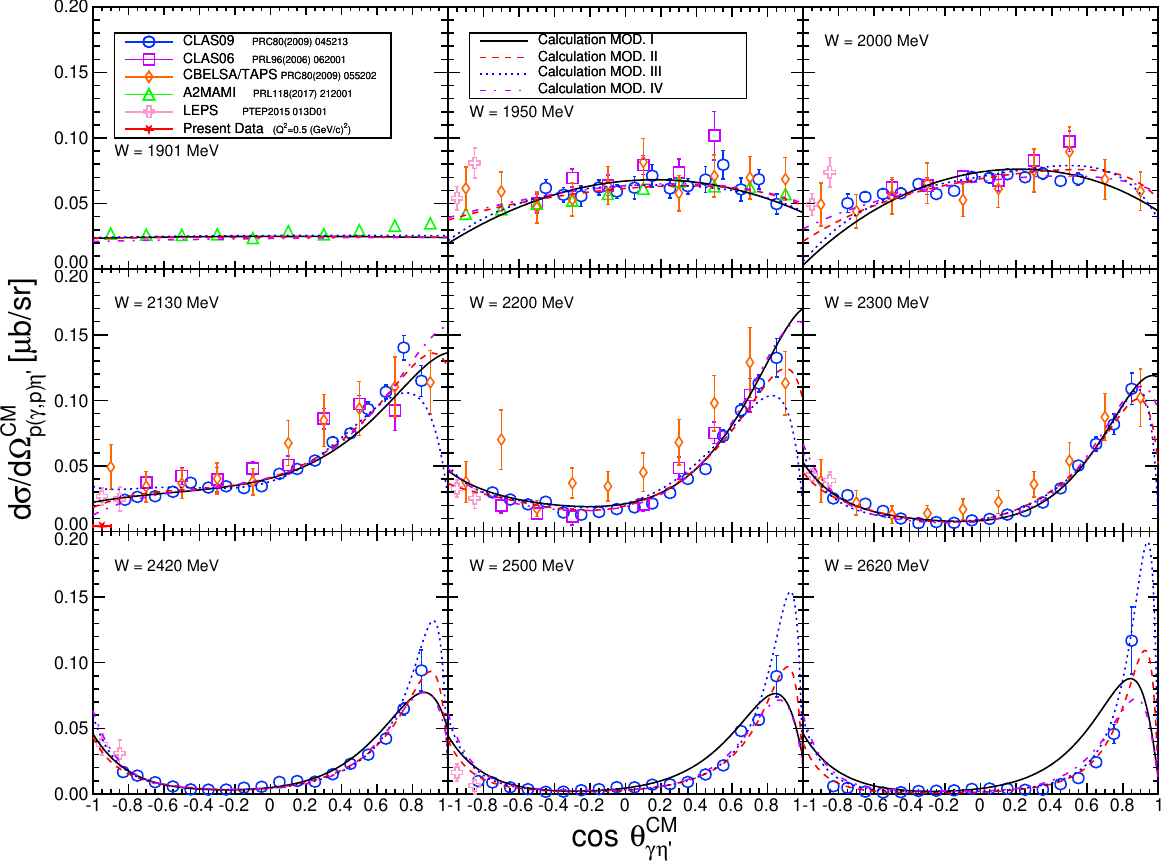}
  \caption{Angular dependence of $\eta '$ photoproduction with various energy range (excerpt). Plots with different colors corresponds to data by different experiment including CLAS06~\cite{CLAS06}, CLAS09~\cite{CLAS09}, CBELSA/TAPS~\cite{CBELSA/TAPS}, A2MAMI~\cite{A2MAMI} and LEPS~\cite{LEPS}. A red star with error bars represents the result of the present result at $W=2130~\mathrm{MeV}$ and $\cos \theta _{\gamma \eta '}^{\mathrm{CM}} \approx -1$. Smooth curves give the solution of Model~I--IV. Note that the value of $W$ indicated in each panel corresponds to the one assumed in the theoretical calculation. For the experimental data, we plot the bin whose central value of $W$ is closest to the assumed value.}
  \label{test}
\end{figure}

In the present study, we examine several more diverse resonance sets by comparing our new calculations with the results of this experiment, whereas in Ref.~\cite{Skoupil} the resonance set that best reproduces the photoproduction data was uniquely determined.
We adopt four different resonance sets, which we label as Model~I, Model~II, Model~III and Model~IV.
Table~\ref{tb:ResonanceList} lists the resonances included in each model.
%table
\begin{table}[btp]
 \caption{Summary of resonance sets incorporated into the isobar model calculation Model~I--IV. All three models have $\rho \left( 770 \right)$ and $\omega \left( 782 \right)$ meson coupling in $t$-channel. The diamond symbols $\diamond$ indicates that the mass and width have been slightly modified from the PDG summary values.}
 \label{tb:ResonanceList}
 \centering
  \begin{tabular}{cc|cccc}
   \hline
   Resonance & PDG status~\cite{PDG} & Model I & Model II & Model III & Model IV\\
   \hline \hline
   $\rho \left( 770 \right)$ & & $\checkmark$ & $\checkmark$ & $\checkmark$ & $\checkmark$ \\
   $\omega \left( 782 \right)$ & & $\checkmark$ & $\checkmark$ & $\checkmark$ & $\checkmark$ \\
   $N \left( 1860 \right)  \frac{5}{2} ^{+}$ & ** &  &  & $\checkmark$ & \\
   $N \left( 1875 \right)  \frac{3}{2} ^{-}$ & *** &  &  &  & $\checkmark$ \\
   $N \left( 1880 \right)  \frac{1}{2} ^{+}$ & *** &  & $\checkmark$ & $\checkmark$ & \\
   $N \left( 1895 \right)  \frac{1}{2} ^{-}$ & **** & $\checkmark$  & $\checkmark$ & $\checkmark$ & $\diamond$ \\
   $N \left( 1900 \right)  \frac{3}{2} ^{+}$ & **** & $\checkmark$  & $\checkmark$ & $\checkmark$ & $\diamond$ \\
   $N \left( 2000 \right)  \frac{5}{2} ^{+}$ & ** & & &  $\checkmark$ & \\
   $N \left( 2060 \right)  \frac{5}{2} ^{-}$ & *** & & $\checkmark$ & & \\
   $N \left( 2100 \right)  \frac{1}{2} ^{+}$ & *** & $\checkmark$ & $\checkmark$ & & $\diamond$ \\
   $N \left( 2120 \right)  \frac{3}{2} ^{-}$ & *** & $\checkmark$ & $\checkmark$ &  $\checkmark$ & $\checkmark$\\
   \hline
  \end{tabular}
\end{table}
Model~I selects the four-star resonances $N\left(1895\right)\frac{1}{2}^{-}$ and $N\left(1900\right)\frac{3}{2}^{+}$ and the three star resonances $N\left(2100\right)\frac{1}{2}^{+}$ and $N\left(2120\right)\frac{3}{2}^{-}$ that have non-zero branching ratio to the $\eta'p$ production channel, based on the summary provided by the PDG~\cite{PDG}.
Model~II adds two additional resonances, $N\left(1880\right)\frac{1}{2}^{+}$ and $N\left(2060\right)\frac{5}{2}^{-}$, both of which are rated with three-star for existance by PDG~\cite{PDG}.
Model~III selects six resonances that show significant contributions to the total cross sections in the EtaMAID2018 model~\cite{EtaMAID2018}, which accounts both cases of $\eta$ and $\eta'$ photoproduction.
Model~IV is identical to Model~B in Ref.~\cite{Skoupil}; it incorporates five resonances of $N\left(1875\right)\frac{3}{2}^{-}$, $N\left(1895\right)\frac{1}{2}^{-}$, $N\left(1900\right)\frac{3}{2}^{+}$, $N\left(2100\right)\frac{1}{2}^{+}$, and $N\left(2120\right)\frac{3}{2}^{-}$, with slightly modification of masses and widths of $N\left( 1895\right)\frac{1}{2}^{-}$, $N\left( 1900\right)\frac{3}{2}^{+}$ and $N\left(2100\right)\frac{1}{2}^{+}$ (details of these modifications are given in TABLE~II of Ref.~\cite{Skoupil}).
Models~I--III were fitted to the previously measured photoproduction differential cross section data (SAPHIR~\cite{SAPHIR}, CLAS~\cite{CLAS06,CLAS09}, CBELSA/TAPS~\cite{CBELSA/TAPS}, A2MAMI~\cite{A2MAMI}, and LEPS~\cite{LEPS}), whereas Model~IV includes the photon beam asymmetry $\Sigma$ data measured by GRAAL~\cite{GRAAL} and CLAS~\cite{CLASPBA} in the fitting procedure in addition.
Fig.~\ref{test} shows the fitting results of Models~I--IV to the photoproduction database.
The horizontal axis represents the production angle in CM frame, while the vertical axis indicates the differential cross section.
From the upper-left to the lower-right panels, total energy $W$ in CM frame increases from $\mathrm{1901~MeV}$ to $\mathrm{2620~MeV}$.
All models reproduce the overall behavior of the angular and energy ($W$) dependencies reasonably well.
However, the predictions at very forward and  backward angles differ each other.
This discrepancy reflects remaining uncertainties in each model, due to the scarcity of experimental data or/and the sensitivity to the choice of the resonance set.
It should be noted that our electroproduction data at $W = \mathrm{2130~MeV}$ was excluded from the fit.\par

In order to compare our models among each other, we use the Akaike Information Criterion (AIC), which rewards descriptive accuracy and penalizes lack of parsimony according to the number of free parameters.
The AIC reads 
\begin{eqnarray}
  \text{AIC} = 2n + \chi^2/\text{n.d.f.},
\end{eqnarray}
where $n$ is the number of free parameters and $\chi^2/\text{n.d.f.}$ is the $\chi^2$ normalized per degree of freedom~\cite{AIC}.
The meaning of this criterion is derived from the comparison of the AIC values of the models with the model having the lowest AIC value representing the best approximating model.
The simplest way of comparing different models and measuring how much better the best approximating model is compared to the next best models is to calculate the Akaike difference of the $i$-th model, $\text{AIC}_i$, and the best model $\text{AIC}_{\text{min}}$
\begin{eqnarray}
  \Delta_i = \text{AIC}_i - \text{AIC}_{\text{min}}.
\end{eqnarray}
This procedure reflects our interest in the relative performance of the models, not their absolute AIC values.
There is still debate on when a model can be considered uninformative, but as a rough guide, models with $\Delta_i$ values less than $2$ are considered to be as good as the best model and models with $\Delta_i$ up to $6$ should not be discounted.
Only above this value, model rejection might be considered~\cite{BA}.
More commonly seen measure is the Akaike weight, $w_i$, which is for a given model, $i$, calculated as
\begin{eqnarray}
  w_i = \frac{\text{exp}(-\Delta_i/2)}{\sum_{r=1}^R \text{exp}(-\Delta_r/2)},
\end{eqnarray}
where $R$ is the number of considered models (here $R=4$).
The Akaike weight is a value between $0$ and $1$, with the sum of Akaike weights of all models in the candidate set being $1$.
It can be considered analogous to the probability that a given model is the best approximating model~\cite{BA}.
In Table~\ref{tab:AIC}, we collect the numbers of free parameters, $n$, values of $\chi^2/\text{n.d.f.}$, and also values of the Akaike Information Criterion (AIC), Akaike differences, $\Delta_i$, and Akaike weights, $w_i$, of all considered models.
It should be noted that, although only Model~IV additionally uses the data of photon beam asymmetry for fitting, the resulting difference in the number of data points ($1044$ for Models~I--III and $1120$ for Model~IV) does not affect the outcome of the AIC analysis.
Comparing our models with AIC, Akaike differences and Akaike weights, we see that the Model~I has the lowest value of AIC and thus can be considered the best approximating model.
Even though the $\chi^2$ value of this model is the largest one among the considered models, it is compensated by having the least number of free parameters.
Moreover, Model~I has a $w_i$ of $0.77$ which can be interpreted as meaning that there is $77~\%$ chance that it really is the best approximating model describing the data given the candidate set of models considered.
Nevertheless, the Model~IV with its moderate number of free parameters and substantially lower value of $\chi^2/\text{n.d.f.}$ and  with its $\Delta_i$ around $3$ can be considered roughly as good as Model~I.
On the other hand, we might consider rejecting Model~III from further considerations based on its very large value of $\Delta_i$ and negligible $w_i$. \par
\begin{table}[h]
    \centering
    \caption{Numbers of free parameters, $n$, values of $\chi^2/\text{n.d.f.}$, Akaike Informatin Criterion (AIC), Akaike differences, $\Delta_i$, and Akaike weights, $w_i$, of models I, II, III, and IV.}
    \begin{tabular}{l r r r r r r} \hline
    Model & $n$ & $\chi^2/\text{n.d.f.}$ & AIC   & $\Delta_i$ & $w_i$ \\ \hline \hline
    I     &  13 &               2.89     & 28.89 &       0.00 &  0.77 \\
    II    &  16 &               1.94     & 33.94 &       5.05 &  0.06 \\
    III   &  17 &               2.51     & 36.51 &       7.62 &  0.02 \\
    IV    &  15 &               2.15     & 32.15 &       3.26 &  0.15 \\ \hline
    \end{tabular}
    \label{tab:AIC}
\end{table}

\subsection{Discussion for $Q^{2}$-dependence}
\begin{figure}[btp]
  \centering
  \includegraphics[width=0.75\linewidth]{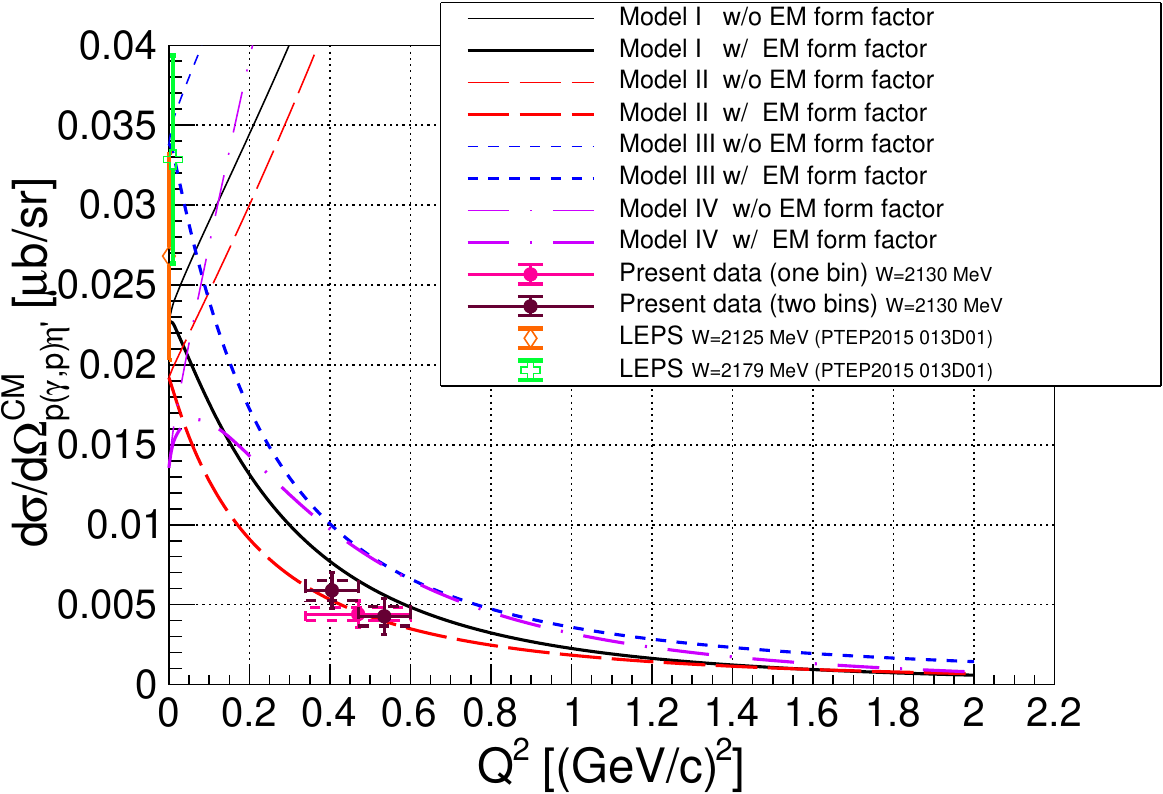}
  \caption{$Q^{2}$-dependence of the differential cross section in electroproduction of $\eta '$ mesons. Bold/thin curves correspond to calculation with/without the electromagnetic formfactor. The pink plot with errors represents the result of the present experiment in a single bin. In contrast, the brown plots are the results divided into two bins. The error bars to each plot represent statistical errors, while the boxes represent systematic errors. The green and orange plots at $Q^{2} = 0$ (orverlapping each other) are the value of existing data at nearby $W$ of photoproduction measured by LEPS~\cite{LEPS}.}
  \label{Q2Dependence}
\end{figure}

Fig.~\ref{Q2Dependence} shows the result of the present experiment overlaid on the $Q^{2}$-dependence plots obtained from Models~I--IV.
The kinematics assumed in the four models follow the experimental conditions: $W = \mathrm{2130~MeV}$, $\cos\theta_{\gamma\eta'}^{\mathrm{CM}} = -0.95$, and $\epsilon = 0.7$.
The thin lines represent calculations without including the electromagnetic form factor, while the bold lines correspond to those with the electromagnetic form factor.
The data point at $Q^{2} = 0.46 \left(\mathrm{GeV/}c\right)^{2}$ corresponds to our measurement; the pink marker represents the result obtained from the full dataset, while the brown markers correspond to results from the dataset divided into two bins.
Error bars indicate statistical uncertainties of the cross section, and boxes represent systematic uncertainties.
The points at $Q^{2} = 0$ are photoproduction data at $W = \mathrm{2125~MeV}$ and $W = \mathrm{2179~MeV}$ at $\cos\theta\approx-1$, reported by LEPS collaboration~\cite{LEPS}.
Our measured cross section is approximately one-sixth of the nearby $W$ photoproduction values.\par

Calculations without the electromagnetic form factor reflect the purely dynamical $Q^{2}$-dependence arising from the transition matrix of meson electroproduction.
These calculations exhibit an rapidly increasing behavior with $Q^{2}$, which is completely inconsistent with our experimental result.
On the other hand, calculations that incorporate the electromagnetic form factor show a quenched behavior as $Q^{2}$ increases.
These models reproduce both data of electroproduction at $Q^{2}>0$ and photoproduction at $Q^{2}=0$ reasonably well, and the observed reduction of the differential cross section is still within the uncertainty arising from the choice of the resonance set assumed in the theoretical calculations.
However, model~IV, which provided the best agreement with both data of differential cross section and photon beam asymmetry in Ref.~\cite{Skoupil}, not only underestimates the LEPS data point at this energy, but also yields a cross section larger than our new data point.
This small discrepancy may possibly originate from an insufficient suppression made by the electromagnetic form factor.
If the electromagnetic form factor produced a faster decrease in the  $Q^{2}$-dependence, model~IV would fit better the new data.
Alternatively, as employed in Ref.~\cite{BS3}, introducing the longitudinal couplings of baryons to the photon may represent another possibility.
By tuning the newly introduced coupling parameters, one could modify the $Q^{2}$-dependence without changing the results for photoproduction cross section.\par

\subsection{Discussion for $W$-dependence}
\begin{figure}
    \centering
    \begin{subfigure}{0.49\textwidth}
        \centering
        \includegraphics[width=\linewidth]{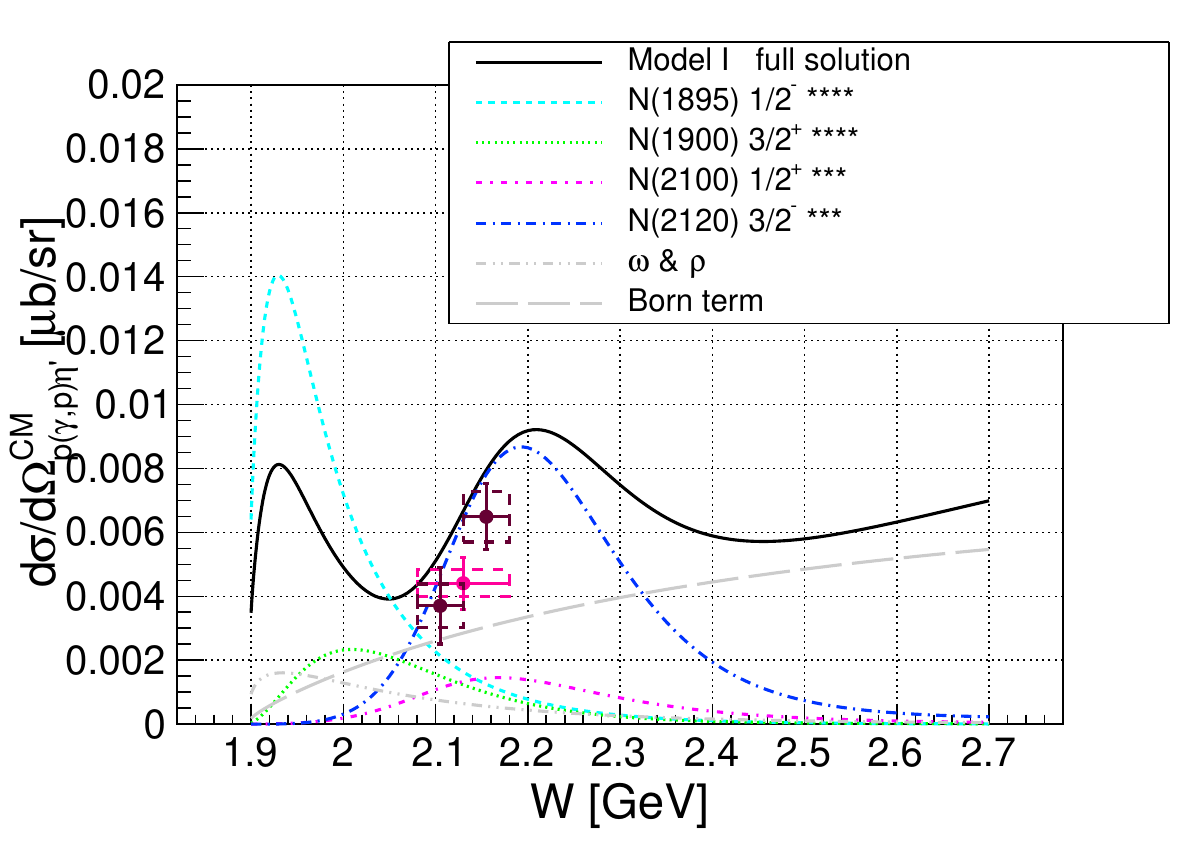}
        \caption{Compared with Model~I.}
        \label{fig:subfig1}
    \end{subfigure}
    \hfill
    \begin{subfigure}{0.49\textwidth}
        \centering
        \includegraphics[width=\linewidth]{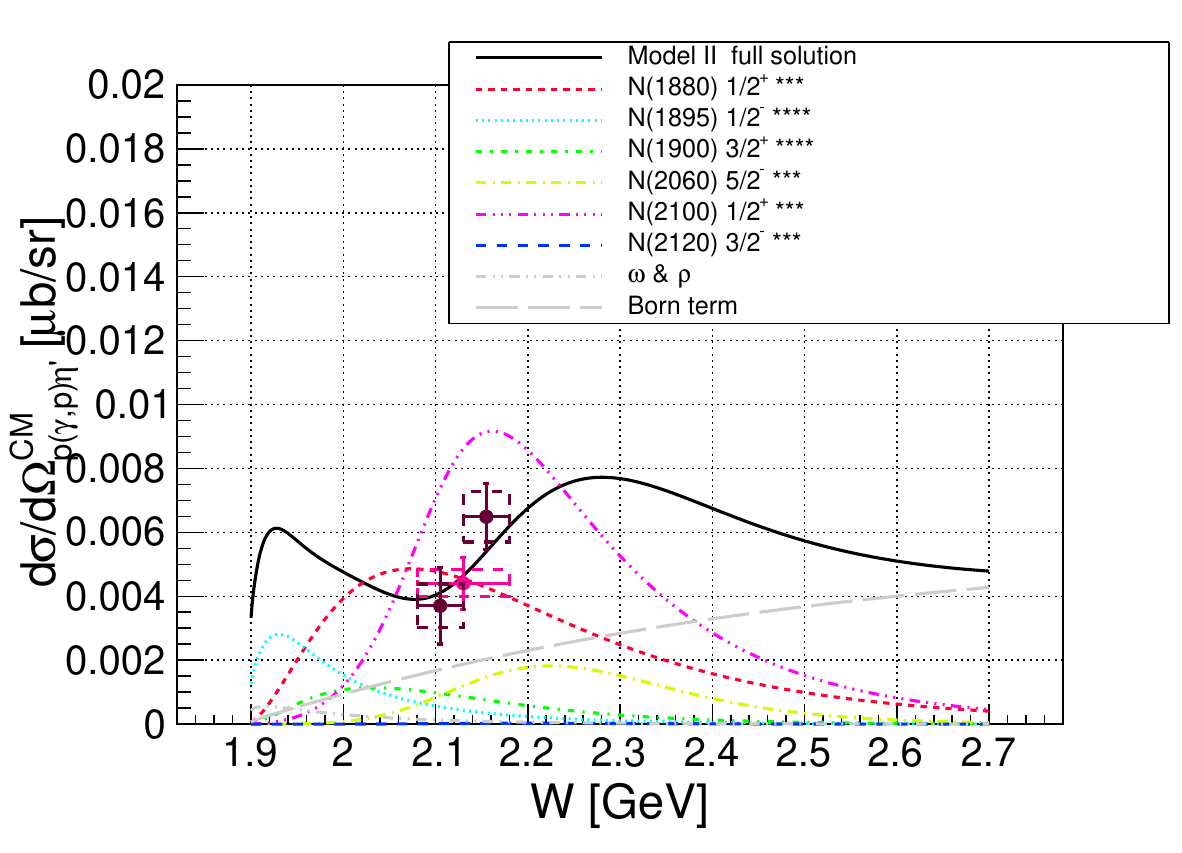}
        \caption{Compared with Model~II.}
        \label{fig:subfig2}
    \end{subfigure}

    \vspace{1em}

    \begin{subfigure}{0.49\textwidth}
        \centering
        \includegraphics[width=\linewidth]{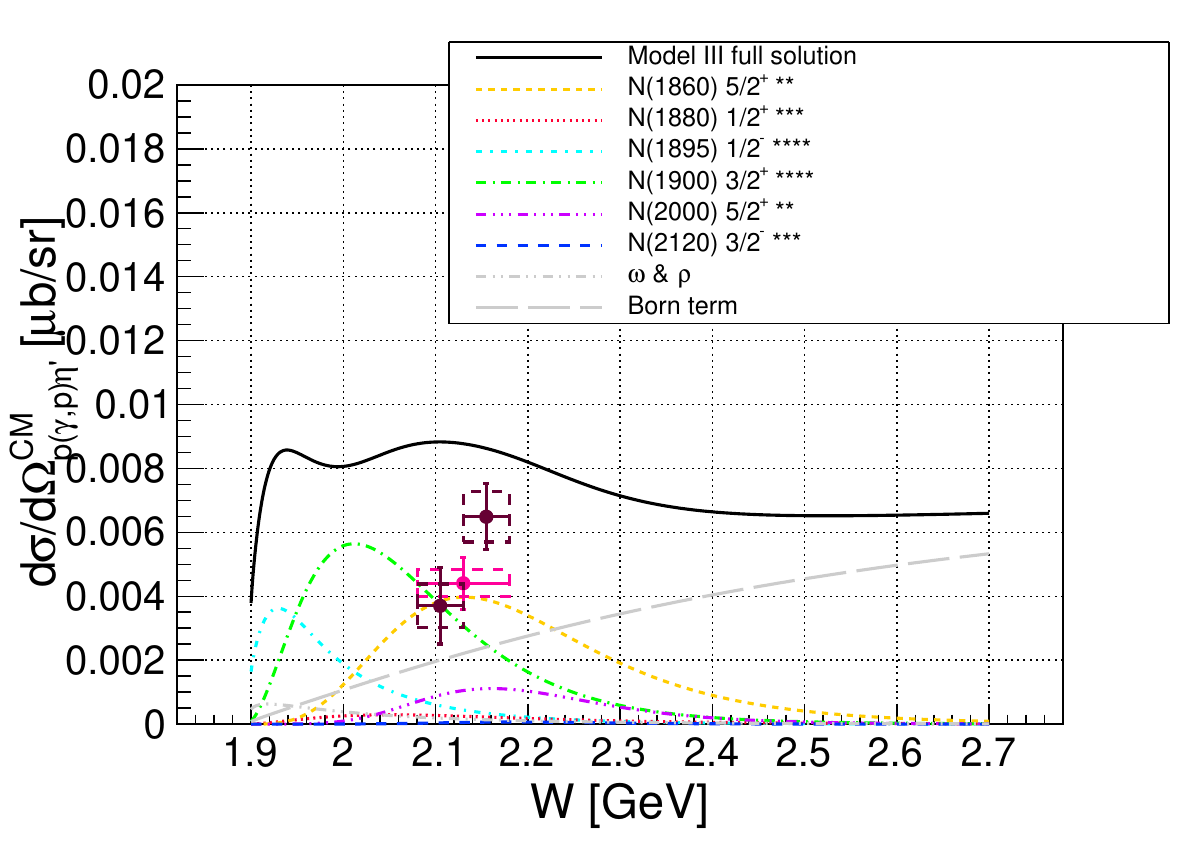}
        \caption{Compared with Model~III.}
        \label{fig:subfig3}
    \end{subfigure}
    \hfill
    \begin{subfigure}{0.49\textwidth}
        \centering
        \includegraphics[width=\linewidth]{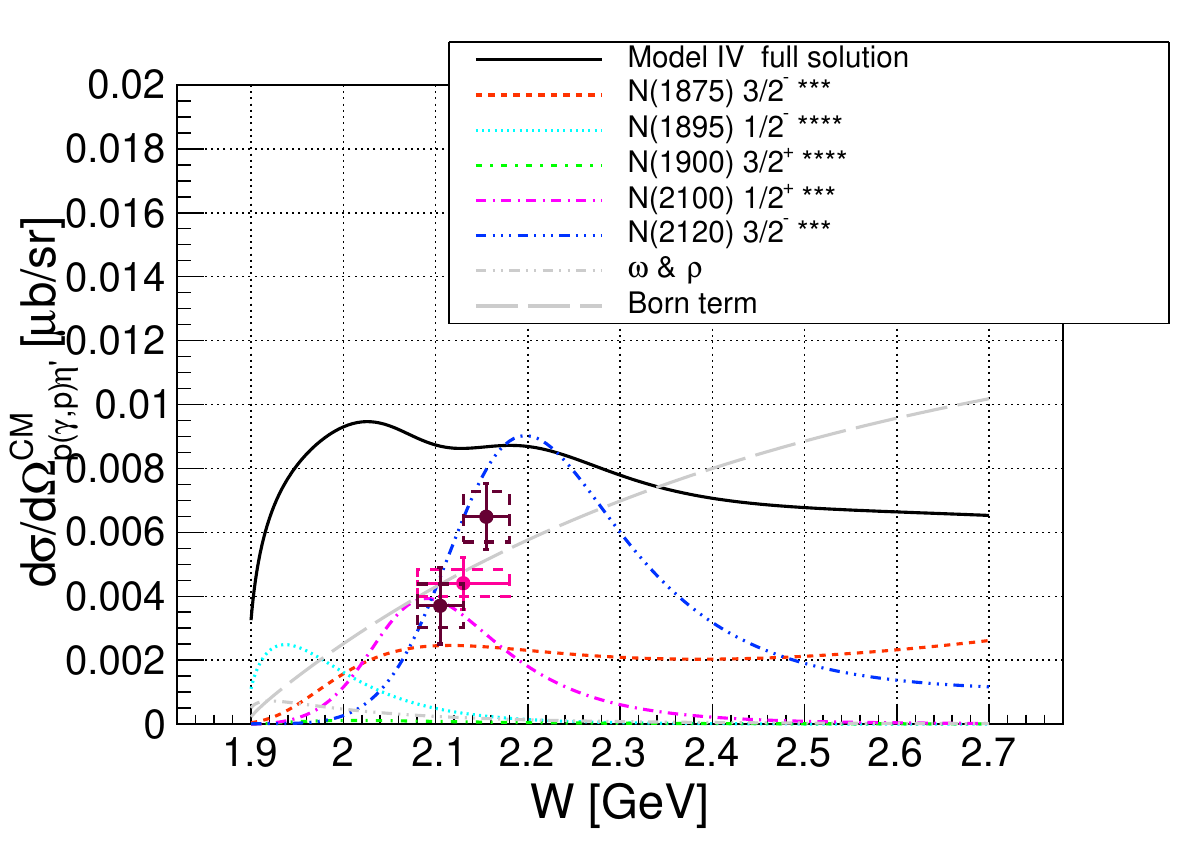}
        \caption{Compared with Model~IV.}
        \label{fig:subfig4}
    \end{subfigure}

    \caption{$W$-dependence of the differential cross section of the $\eta '$ electroproduction. The pink plot with errors represents the result of the present experiment in a single bin. In contrast, The brown plots are the results divided into two bins. The error bars to each plot represent statistical errors, while the boxes represent systematic errors. The solid curves give the full solutions of Model~I--IV at the specific kinematics, and the other curves give the individual contributions of different resonances.}
    \label{fig:combined}
\end{figure}

Fig.~\ref{fig:combined} compare the experimental results of this study with the $W$-dependence predicted by Models~I--IV.
As before, the pink markers represent the results from the full dataset, while the brown markers correspond to data divided into two bins.
All four calculations assume the kinematic conditions of $Q^{2}=0.46~\left(\mathrm{GeV/}c\right)^{2}$, $\cos\theta_{\gamma\eta'}^{\mathrm{CM}}=-0.95$, and $\epsilon=0.7$.
The solid curves give the full solutions of Model~I--IV.
The other curves in various colors indicate the individual contributions from each resonance.\par

The $W$-dependence predicted by the four theoretical models differs significantly.
The contribution from each resonance is determined to best reproduce the angular dependence of the photoproduction data; however, which resonance contributes most strongly under the given kinematics depends on the assumed resonance set.
This discrepancy is particularly apparent in the $W$-dependence of backward-angle production.
Among the four models, Model~II is the most preferable in describing the present experimental data.
Furthermore, when comparing the rise and fall behavior of the cross sections near the energy region of interest, Models~I and II increase with $W$, while Models~III and IV exhibit an almost flat.
Looking into the contributions from individual resonances, Model~I is dominated by $N\left(2130\right)\frac{3}{2}^{-}$, while Model~II is primarily driven by $N\left(2100\right)\frac{1}{2}^{+}$.
In contrast, Model~III attributes the cross section mainly to couplings with lower-energy resonances.
In Model~IV, the newly introduced $N\left(1875\right)\frac{3}{2}^{-}$ contributes almost as a constant term in the region above $W > \mathrm{2~GeV}$.
Another noteworthy feature is that the Born-term contribution is larger compared to the other three calculations.
The experimental results obtained in this study show an increasing trend beyond $W=\mathrm{2130~MeV}$, prefering Model~I and Model~II.
Therefore, the present measurement suggests a significant role played by one or more resonances near $W\sim\mathrm{2100~MeV}$.
This new data is expected to provide a new constraint for future resonance studies; however, to enable a more detailed discussion, experimental data covering a wider $W$ range or data with improved statistical precision will be required.\par

\section{Summary}
The electroproduction of $\eta '$ mesons from a $\mathrm{^{1}H}$ target was analyzed from calibration data of the $\Lambda$ hypernuclear spectroscopy experiment at Jefferson Lab.
A peak with a mass of $m_{\eta'}=\mathrm{958~MeV/}c^{2}$ was observed in the missing mass spectrum of the $\mathrm{^{1}H}\left( e,e'p \right) X$ reaction.
The observed number of events of $\eta'$ mesons, when converted to the differential cross section of the $\gamma ^{*} +p \rightarrow \eta ' +p $ reaction at $W = \mathrm{2130~MeV}$, $Q^{2} = 0.46~\left( \mathrm{GeV/}c \right) ^{2}$, and $\cos \theta ^{\mathrm{CM}}_{\gamma \eta'} \approx -1.0$ in the One-Photon-Exchange Approximation, corresponds to $\left( \frac{\mathrm{d}\sigma _{\gamma^{*}p\rightarrow \eta 'p}}{\mathrm{d}\Omega_{\eta '}} \right) ^{\mathrm{CM}} = 4.4 \pm 0.8 ~\left( \mathrm{stat.} \right) \pm 0.4 ~\left( \mathrm{sys.} \right)~\left[ \mathrm{nb/sr}\right]$.
This value is approximately one-sixth of that of backward photoproduction at $Q^{2} = 0$ measured by the LEPS collaboration.
To analyze the present experimental results, a theoretical calculation based on the isobar model was demonstrated.
It describes both the database of angular dependence of photoproduction in various past experiments and the virtual photoproduction in the present experiment within the same framework.
The observed decrease in measured differential cross section at finite $Q^{2}>0$ is consistent with the new calculations with an electromagnetic form factor at the virtual-photocoupling vertex.
The $W$-dependence derived by dividing the experimental data into two bins showed a rising behavior at the boundary of $W=2130~\mathrm{MeV}$, and the degree of agreement with the theoretical calculations varies depending on the combination of resonance sets introduced in each theoretical model.
The new data suggests that (a) $N^{*}$ resonance(s) with energy of approximately $\mathrm{2100~MeV}$ might play an important role for the coupling to the $\eta' p$ final state, and imposes new constraints for a future resonance search.

\section*{Acknowledgment}
We thank the Jefferson Lab staff in the Divisions of Physics, Division of Accelerator, and Division of Engineering for their support in carrying out the experiment.
We also gratefully acknowledge the exceptional contribution of the Jefferson Lab target group for designing and safely operating the cryogenic gas target system used in this work.
This work was supported by U.S. Department of Energy (DOE) grant DE-AC05-06OR23177, under which Jefferson Science Associates LLC operates the Thomas Jefferson National Accelerator Facility.
Support for the Argonne National Laboratory group was provided by DOE grant DE-AC02-06CH11357.
This work was supported by the Grant Agency of the Czech Republic under Grant No. 25-18335S.
The Kent State University contribution was supported by U.S. National Science Foundation grant PHY-1714809.
The hypernuclear program at Jefferson Lab is supported by DOE grant DE-FG02-97ER41047.
This work was partially supported by the Grant-in-Aid for Scientific Research on Innovative Areas, ``Toward new frontiers: Encounter and synergy of state-of-the-art astronomical detectors and exotic quantum beams.''
Additional support was provided by JSPS KAKENHI grants Nos.~17H01121, 18H05459, 18H05457, 18H01219, 19J22055, 18H01220, and 24H00219.
This work was also supported by SPIRITS 2020 of Kyoto University, and by the Graduate Program on Physics for the Universe, Tohoku University (GP-PU).\par

% can use a bibliography generated by BibTeX as a .bbl file
% BibTeX documentation can be easily obtained at:
% http://www.ctan.org/tex-archive/biblio/bibtex/contrib/doc/

%\bibliographystyle{ptephy}
%\bibliography{sample}
%
% once the .bbl file has been generated then place the text in your article.

\vspace{0.2cm}
\noindent
%For references,  note how to include DOI information from examples below. 

%This is added by T. Yoneya (editor-in-chief) on 2020/07/09.

\let\doi\relax

%without this code before the command "\begin{thebibliography}{}" , an error will be %flagged. When the bibliography is provided as separate .bib file, then this code %should be placed above the commands "\bibliographystyle{}" and "\bibliography{}" %inside the main TeX file. 

\begin{thebibliography}{99}

\bibitem{GM1}
M.~Gell-Mann, Phys. Rev. \textbf{92}, 833 (1953).

\bibitem{GM2}
M.~Gell-Mann, Phys. Rev. \textbf{125}, 1067 (1962).

\bibitem{GM3}
M.~Gell-Mann, Phys. Lett. \textbf{8}, 214 (1964).

\bibitem{Zweig1}
G.~Zweig, CERN Report No. \textbf{8182}/TH.~401 (1964).
                                      
\bibitem{Zweig2}                      
G.~Zweig, CERN Report No. \textbf{8419}/TH.~412 (1964).

\bibitem{PDG}
Particle Data Group, Phys. Rev. D \textbf{110}, 030001 (2024).

\bibitem{ua1}
G.~t' Hooft, Phys. Rev. D \textbf{14}, 3432 (1976).

\bibitem{ua2}
E.~Witten, Nucl. Phys. B \textbf{156}, 269--283 (1979).

\bibitem{ua3}
G.~Veneziano, Nucl. Phys. B \textbf{159}, 213--224 (1979).

\bibitem{ua4}
D.~Jido \textit{et~al.,} Nucl. Phys. A \textbf{914}, 354--359 (2013).

\bibitem{EMN1}
P.~Costa  \textit{et.~al.,} Phys. Rev. D \textbf{71}, 116002 (2005).

\bibitem{EMN2}
H.~Nagahiro \textit{et.~al.,} Phys. Rev. C \textbf{74}, 045203 (2006).

\bibitem{EMN3}
H.~Nagahiro and S.~Hirenzaki, Phys. Rev. Lett. \textbf{94}, 232503 (2005).

\bibitem{EMN4}
S.~Sakai and D.~Jido, Phys. Rev. Lett. \textbf{88}, 064906 (2013).

\bibitem{TanakaGSI}
Y.~N.~Tanaka \textit{et~al.,} Phys. Rev. C \textbf{97}, 015202 (2018).

\bibitem{TomidaLEPS}
N.~Tomida \textit{et~al.,} Phys. Rev. Lett. \textbf{124}, 202501 (2020).

\bibitem{SAPHIR}
R.~Pl\"{o}tzke \textit{et~al.,} Phys. Lett. B \textbf{444}, 555--562 (1998).

\bibitem{CLAS06}
M.~Dugger \textit{et~al.,} Phys. Rev. Lett. \textbf{96}, 062001 (2006).

\bibitem{CLAS09}
M.~Williams \textit{et~al.,} Phys. Rev. C \textbf{80}, 045213 (2009).

\bibitem{CBELSA/TAPS}
V.~Crede \textit{et~al.,} Phys. Rev. C \textbf{80}, 055202 (2009).

\bibitem{A2MAMI}
V.~Kashevarov \textit{et~al.,} Phys. Rev. Lett. \textbf{118}, 212001 (2017).

\bibitem{LEPS}
Y.~Morino \textit{et~al.,} Prog. Theor. Exp. Phys. \textbf{2015}, 013D01.

\bibitem{Miyoshi}
T.~Miyoshi \textit{et~al.,} Phys. Rev. Lett. \textbf{90}, 232502 (2003).

\bibitem{Iodice}
M.~Iodice \textit{et~al.,} Phys. Rev. Lett. \textbf{99}, 052501 (2007).

\bibitem{Cusanno}
F.~Cusanno \textit{et~al.,} Phys. Rev. Lett. \textbf{103}, 202501 (2009).

\bibitem{Nue}
S.~N.~Nakamura \textit{et~al.,} Phys. Rev. Lett. \textbf{110}, 012502 (2013).

\bibitem{Tang}
L.~Tang \textit{et~al.,} Phys. Rev. C \textbf{90}, 034320 (2014).

\bibitem{Urciuoli}
G.~M.~Urciuoli \textit{et~al.,} Phys. Rev. C. \textbf{91}, 034308 (2015).

\bibitem{Gogami}
T.~Gogami \textit{et~al.,} Phys. Rev. C \textbf{93}, 034314 (2016).

\bibitem{Suzuki}
K.~N.~Suzuki \textit{et~al.,} Prog. Theor. Exp. Phys. \textbf{2022}, 013D01.

\bibitem{Pandey}
B.~Pandey \textit{et~al.,} Phys. Rev. C \textbf{105}, L051001 (2022).

\bibitem{Okuyama}
K.~Okuyama \textit{et~al.,} Phys. Rev. C \textbf{110}, 025203 (2024).

\bibitem{BS12}
D.~Skoupil and P.~Byd\v{z}ovsk\'y, Phys. Rev. C \textbf{93}, 025204 (2016).

\bibitem{BS3}
D.~Skoupil and P.~Byd\v{z}ovsk\'y, Phys. Rev. C \textbf{97}, 025202 (2018).

\bibitem{Skoupil}
D.~Skoupil \textit{et~al.,}, Phys. Rev. C \textbf{112}, 025202 (2025).

\bibitem{MesonBook}
E.~Amaldi \textit{et~al.,} \textit{``Pion-Electroproduction: Electroproduction at Low Energy and Hadron Form Factors,''} Springer Tracts in Modern Physics Vol.~\textbf{83}, Springer-Verlag Berlin Heidelberg (1979).

\bibitem{CEBAF}
F.~Pilat \textit{et.~al.,} Proceedings of LINAC2012, Tel-Aviv, Israel, TH3A02 792--796 (2012).

\bibitem{arc}
S.~N.~Santiesteban \textit{et.~al.,} arXiv:2110.06281.

\bibitem{BCM}
J.-C.~Denard \textit{et.~al.,} Proceedings of the 2001 Particle Accelerator Conference, 2136.

\bibitem{BPM}
P.~Zhu \textit{et.~al.,} Nucl. Instrum. Methods in Phys. Res. A \textbf{808}, 1 (2016).

\bibitem{Target}
S. N. Santiesteban \textit{et.~al.,} Nucl. Instrum. Methods in Phys. Res. A \textbf{940}, 351 (2019).

\bibitem{HallA}
J.~Alcorn \textit{et.~al.,} Nucl. Instrum. Methods in Phys. Res. A \textbf{522}, 294 (2004).

\bibitem{VDC}
K.~G~Fissum \textit{et.~al.,} Nucl. Instrum. Methods in Phys. Res. A \textbf{474}, 108 (2001).

\bibitem{SIMC1}
https://github.com/JeffersonLab/simc\_gfortran

\bibitem{SIMC2}
A.~Uzzle, Doctoral thesis, Hampton University (2002).

\bibitem{EVMD1}
M.~F.~Gari and W.~Kr\"{u}mpelmann, Phys. Rev. D \textbf{45}, 1817 (1992).

\bibitem{EVMD2}
M.~F.~Gari and W.~Kr\"{u}mpelmann, Phys. Rev. D \textbf{46}, 484 (1992).

\bibitem{EMFF1}
R.~A.~Williams \textit{et al.,} Phys. Rev. C \textbf{46}, 1617 (1992).

\bibitem{EMFF2}
E.~L.~Lomon, Phys. Rev. C \textbf{66}, 045501 (2002).

\bibitem{EtaMAID2018}
L.~Tiator \textit{et al.,} Eur. Phys. J. A \textbf{54}, 210 (2018).

\bibitem{GRAAL}
P.~Levi Sandri \textit{et al.,} Eur. Phys. J. A \textbf{51}, 77 (2015).

\bibitem{CLASPBA}
P.~Collins \textit{et al.,} Phys. Lett. B \textbf{771}, 213--221 (2017).

\bibitem{AIC}
P.~Byd\v{z}ovsk\'{y} \textit{et~al.,} Phys. Rev. C \textbf{104}, 065202 (2021).

\bibitem{BA}
K.~P.~Burnham and D.~R.~Anderson, \textit{``Model Selection and Multimodel Inference: A Practiral Information---Theoretic Approach, Second Edition,''} Springer New York (2002).

\end{thebibliography}

%\appendix
%\section{Appendix head}
%This is the sample text. This is the sample text. This is the sample text. This is the sample text.

\end{document}